\documentclass[acmsmall,screen]{acmart}

\AtBeginDocument{%
	\providecommand\BibTeX{{%
			\normalfont B\kern-0.5em{\scshape i\kern-0.25em b}\kern-0.8em\TeX}}}

\usepackage{amsmath}
\usepackage{graphicx}
\usepackage[T1]{fontenc}   
\usepackage{ccicons}          
\usepackage{paralist}
\usepackage{hyperref}
\usepackage[nolist]{acronym}
\usepackage{color}

 \usepackage{multirow}
 \usepackage{colortbl}
 \usepackage{hhline}

\def\hb{\hbox to 10.7 cm{}}

\newcommand{\quotes}[1]{``#1''}
\newtheorem{hypothesis}{Hypothesis}

\newcommand{\edit}[1]{#1}

\begin{document}
	
	\title{Generating User-Centred Explanations via Illocutionary Question Answering: From Philosophy to Interfaces}
	
	\author{Francesco Sovrano}
	\email{francesco.sovrano2@unibo.it}
	\orcid{0000-0002-6285-1041}
	\affiliation{%
		\institution{Università di Bologna}
		\city{Bologna}
		\country{Italy}}
	
	\author{Fabio Vitali}
	\email{fabio.vitali@unibo.it}
	\orcid{0000-0002-7562-5203}
	\affiliation{%
		\institution{Università di Bologna}
		\city{Bologna}
		\country{Italy}}
	
	\renewcommand{\shortauthors}{Sovrano, et al.}
	\renewcommand{\shorttitle}{Generating Explanations via Illocutionary Question Answering}
	
	
	\begin{abstract}
		We propose a new method for generating explanations with \ac{AI} and a tool to test its expressive power within a user interface.
		In order to bridge the gap between philosophy and human-computer interfaces, we show a new approach for the generation of interactive explanations based on a sophisticated pipeline of \ac{AI} algorithms for structuring natural language documents into knowledge graphs, answering questions effectively and satisfactorily. 
		With this work we aim to prove that the philosophical theory of explanations presented by Achinstein can be actually adapted for being implemented into a concrete software application, as an interactive and illocutionary process of answering questions.
		Specifically, our contribution is an approach to frame \textit{illocution} in a computer-friendly way, to achieve user-centrality with statistical question answering.
		\edit{Indeed, we frame the \textit{illocution} of an explanatory process as that mechanism responsible for anticipating the needs of the explainee in the form of unposed, implicit, archetypal questions, hence improving the user-centrality of the underlying explanatory process.
		Therefore, we hypothesise that if an explanatory process is an illocutionary act of providing content-giving answers to questions, and illocution is as we defined it, the more explicit and implicit questions can be answered by an explanatory tool, the more usable (as per ISO 9241-210) its explanations.}
		We tested our hypothesis with a user-study involving more than 60 participants, on two XAI-based systems, one for credit approval (finance) and one for heart disease prediction (healthcare).
		The results showed that increasing the \textit{illocutionary power} of an explanatory tool can produce statistically \textit{significant} improvements (hence with a p-value lower than 0.05) on effectiveness. 
		This, combined with a visible alignment between the increments in effectiveness and satisfaction, suggests that our understanding of \textit{illocution} can be correct, giving evidence in favour of our theory.
		
	\end{abstract}

	

	\maketitle
	
	\section{Introduction} \label{sec:introduction}
	The complexity of modern software and the increasing discomfort of humans towards the correctness and fairness of the output of such complex systems has caused the birth and growth of a new discipline to reduce the distance between individuals, society, and machines: \ac{XAI}.
	
	Governments have also started to act towards the establishment of ground rules of behaviour from complex systems, for instance through the enactment of the European \acf{GDPR} (2016\footnote{Regulation (EU) 2016/679.}), which identifies \emph{fairness}, \emph{lawfulness}, and in particular \emph{transparency} as basic principles for every data processing tools handling personal data; even creating a new \emph{Right to Explanation} for individuals whose legal status is affected by a solely-automated decision. 
	
	By and large, literature agrees that explanations in \ac{XAI} systems are answers to a question, usually about the outcome of a computation. 
	For some time, the question was expected to be focusing on the individual computation performed by the system (a \textit{local} question) and to the \textit{causes} of such outcome, so it could be phrased as a \quotes{why} or \quotes{how} question, and specifically a \quotes{Why did I obtain this result (as opposed to some other ones)?}. 
	Over time, more and more sophisticated expectations arose about which questions could be identified as explanation requests, and whether the explanation provided would be the same for all requests, or just one out of a family within which to choose via an abductive process (i.e., the \quotes{best one} of many possible answers) to achieve \textit{user-centrality}.
	
	Merely getting access to the outcome and the internal state of a complex \ac{AI} computation is important but not sufficient to handle the variety of different explanatory goals that we expect to find in our users. 
	That is to say, \ac{XAI} systems alone do not provide sufficient information to answer to all our archetypal questions, but, rather, their output must be somehow reorganised and enriched with additional information, both local and global, i.e., about and beyond the scope and the specificity of the individual computation of the process.
	
	This is why we are interested in designing and developing software for generating user-centred explanations, in the attempt to shed more light on the difference it bears in terms of effectiveness with respect to non-pragmatic approaches. 
	More precisely, we want to understand how to structure information in order to facilitate the production of pragmatic explanations of complex decision-making processes.
	
	We acknowledge that we are not the first to try to model an explanatory process. In literature there were various efforts in this direction and a long history of debates and philosophical traditions, often rooted in Aristotle's works and those of other philosophers. 
	Among the many philosophical theories proposed over the last few centuries some are now considered fallacious, albeit historically useful (i.e. Hempel's \cite{hempel1965aspects}).
	
	\edit{In this paper, we propose a new approach to explanations in Artificial Intelligence, extending \cite{sovrano2021philosophy}. 
	Our own approach is based on Achinstein's theory of explanations (\citeyear{achinstein1983nature}) \cite{achinstein2010evidence}, where explanations are the result of an \textit{illocutionary act} of answering to a question. 
	In particular, it means that there is a subtle and important difference between simply \quotes{answering to questions} and \quotes{explaining}, and this difference is \textit{illocution}: a deliberate intent of producing the \quotes{conventional consequences} of the act \cite{austin1975things}, that in the case of explaining are \textit{understanding}, while in the case of promises are \textit{commitment}, etc..
	For example, answering \quotes{I am fine} to the question \quotes{How are you doing?} is not an explanation, but answering \quotes{I am fine because I was worried I could have tested positive to COVID-19, but I am not and etc..} sounds more like an explanation because of the intent to produce an understanding about \quotes{how I am}.}
	
	In this sense, questions are the main mechanism for an explainee to express her/his own needs, favouring the user-centrality of explanations.
	Some questions may be explicit and others not, some may lose importance over time or vice versa, but normally a user is fully satisfied with explanations only when they effectively convey full coverage of relevant answers for all of his or her goals of understanding.
	Though, modelling an explanatory process as a standard \ac{QA} process gave us the first impression of being a little bit unrealistic. 
	
	Think of the following example of the \quotes{university lectures}: students (the explainees) follow the lessons to acquire (initially obscure) information provided by the professor (the explainer). A lesson can normally include the intervention of students in the form of observations and/or questions, but these interventions are, in practice, always after an initial phase of information acquisition.
	In other terms, the initial overview given by the professor may not be the answer to any preliminary question, especially if the students know absolutely nothing about what the professor is supposed to say. 
	Regardless this apparent lack of a question, we might all agree that the professor could actually explain something good to the students.
	
	At this point it would seem that Achinstein's theory, being based on question-answering, fails to capture the need for preliminary overviews during an explanatory process, as in the \quotes{university lectures} example. 
	Despite this first impression, we think that overviews can be generated as answers as well, therefore partially confirming Achinstein's original theory. 
	
	In fact, for the generation of an overview it is necessary (for the professor) to select and group information appropriately, so as to facilitate the production of different explanatory paths for different users (the students), and the way these clusters of information are created is by anticipating and answering implicit and archetypal questions (e.g. Why X? What is X for? How is X? When was X? etc..). 	
	In particular, we leverage on a subtle and important difference between \quotes{answering to questions} and \quotes{explaining}: \textit{illocution}. 
	
	\edit{According to Achinstein, explaining is when \quotes{S utters u with the intention that his utterance of u renders q understandable by producing the knowledge, of the proposition expressed by u, that it is a correct answer to Q} \cite{achinstein2010evidence}.
	The problem with this philosophical definition of \textit{illocution} is that it is too abstract to be implementable in a software, requiring to concretely find a way to formally frame what is a deliberate intent of explaining.
	This is why we propose a more precise and computer-friendly denotation of \textit{illocution} in this context, as the act of pertinently and deliberately answering to \textit{implicit} (i.e. archetypal) \textit{questions} characterised by the user.}
	
	\edit{In other terms, we depart from Achinstein's definition, asserting that \textit{illocution} is the main mechanism responsible for anticipating unposed or implicit questions/goals, shaping the underlying explanatory process as \textit{more user-centred}, helping both the explainee and the explainer in consuming less resources while communicating, reducing the amount of explanatory steps.
	More precisely, we hypothesise that given an arbitrary explanatory process, increasing its ability to answer both explicit and implicit questions results in the generation of more usable (as per ISO 9241-210) explanations.
	So, if explaining is indeed an illocutionary act of question answering and illocution in explaining is (as previously defined)  a deliberate intent of producing new \textit{understandings} in an explainee by answering also to unposed/implicit questions, then the more an explanatory process is implemented as an illocutionary act of producing content-giving answers to questions, the more it can meet the explanatory goals of a user, the more it is going to be usable (as per ISO 9241-210).
	In fact, a good degree of usability is usually achieved when the specific needs of a user are met by the (explanatory) system.} 
	
	Hence, we designed a novel pipeline of AI algorithms for the generation of pragmatic explanations through the extraction and structuration of an \ac{ES} \cite{sovrano2020modelling}, intended as all the possible explanations (about something to be explained) reachable by a user through an explanatory process, via a pre-defined set of actions, i.e. \textit{Open Question Answering} and \textit{Overviewing}.
	This pipeline is meant to organize the information contained in non-structured documents written in natural language (e.g. web pages, pdf, etc..), allowing efficient information clustering, according to a pre-defined set of archetypal questions.
	
	To verify our hypothesis and evaluate our algorithm, we ran a user-study to compare the usability of the explanations generated through our novel pipeline against classical, one-size-fits-all, static XAI-based explanatory systems.
	\edit{The experiment consisted in explaining to more than 60 unique participants a credit approval system (based on a simple Artificial Neural Network and on CEM\cite{dhurandhar2018explanations}) and an heart disease predictor (based on XGBoost\cite{chen2016xgboost} and TreeShap\cite{lundberg2020local}) in different ways, with different degrees of \textit{illocutionary power} and different mechanisms for the user to ask their own questions explicitly.}
	
	More in detail, to understand the validity of our hypothesis, we compare three different explanatory approaches. 
	\edit{The first approach (Overwhelming Static Explainer; OSE in short) is fully static, dumping on the user complex amounts of information, without any re-elaboration or explicit attempt to answer (implicit or not) questions.
	While the second (How-Why Narrator; HWN in short) and the third (Explanatory AI for Humans; YAI4Hu in short) approach are an interactive version of the first one and they are based on our proposed pipeline.}
	
	\edit{HWN is specifically designed to answer exclusively to \quotes{how} and \quotes{why} archetypal questions, not allowing the users to explicit their own questions.
	On the other side YAI4Hu is designed to have a much greater illocutionary power, answering also to implicit \quotes{what} questions and many others, and (differently from the other systems) it empowers the users with the ability to ask their own questions.
	These tools were designed so that comparing their usability scores would indirectly allow us to isolate and measure the effects of illocution, implicit and explicit question answering, in the generation of user-centred explanations.}
	
	The experiment results gave us enough statistical insights to conclude that increasing the illocutionary power of an explanatory process, and its ability to answer the explicit questions of an explainee, have the potential to produce a statistically significant improvement (hence with a p-value lower than 0.05) on effectiveness. This, combined with a visible alignment between the increments in effectiveness and satisfaction, suggest that our understanding of \textit{illocution} can be correct, favouring the usability of an explanatory process.
	
	This paper is structured as follows. 
	In Section \ref{sec:background} we provide a brief introduction to the contemporary philosophical developments in the theory of explanations, focusing on Achinstein's, and discussing how and in what measure it is aligned to state-of-the-art, especially with respect to \ac{XAI}.
	In Section \ref{sec:proposed_solution} we describe our proposed solution, inspired by Achinstein's, going through the details of what is \textit{illocution} for us and how to achieve it, following in Section \ref{sec:proof_of_concept} detailed instructions of how to implement a proof of concept algorithm for user-centred explanations via illocutionary question answering.
	\edit{In Section \ref{sec:experiments} we present a few experiments to validate the proposed solution, evaluating the proof of concept with a user-study on two XAI-based systems for credit approval (finance) and heart disease prediction (healthcare) explained through three different approaches to explanations.} 
	Finally, in Section \ref{sec:results} we show and discuss the obtained results, drawing the conclusions in Section \ref{sec:conclusions}.
	
	\section{Background and Related Work} \label{sec:related_work} \label{sec:background}
	In this Section we will briefly introduce Achinstein's Theory of Explanations and discuss its alignment with the existing literature on \ac{XAI}.
	
	\subsection{Achinstein's Theory of Explanations} \label{sec:achinstein}
	Being able to automatically generate explanations has attracted the interest of the scientific community for long.
	This interest has increased together with the importance of AI in our society and the growing need to explicate the complexity of modern software systems.
	
	Understanding what constitutes an explanation is a long-standing problem, with a complex history of debates and philosophical traditions, often rooted in Aristotle's works and those of other philosophers.
	According to \citeauthor{mayes2005theories} \cite{mayes2005theories}, explanation in philosophy has been conceived within the following five traditions: 
	\begin{itemize}
		\item \textbf{Causal Realism} \cite{salmon1984scientific}: explanation as a non-pragmatic articulation of the fundamental causal mechanisms of a phenomenon.
		\item \textbf{Constructive Empiricism} \cite{van1980scientific}: epistemic theory of explanation that draws on the logic of why-questions and on a Bayesian interpretation of probability.
		\item \textbf{Ordinary Language Philosophy} \cite{achinstein1983nature}: the act of explanation as the \textit{illocutionary} attempt to produce understanding in another by answering questions in a pragmatic way.
		\item \textbf{Cognitive Science} \cite{holland1989induction}: explaining as a process of belief revision, etc..
		\item \textbf{Naturalism and Scientific Realism} \cite{sellars1963philosophy}: rejects any kind of explanation of natural phenomena that makes essential reference to unnatural phenomena. Explanation is not something that occurs on the basis of pre-confirmed truths. Rather, successful explanation is actually part of the process of confirmation itself. 
	\end{itemize}
	What is in common to all these traditions is that all, but the first, are pragmatic, framing explanations as an artefact which effectiveness may change across different explainees.
	
	In 1983, \citeauthor{achinstein1983nature} was one of the first scholars to analyse the process of generating explanations as a whole, introducing his philosophical model of a \textit{pragmatic} explanatory process.
	
	According to the model, explaining is an \textit{illocutionary} act coming from a clear intention of producing new understandings in an explainee by providing a correct content-giving answer to an open question. 
	Therefore, according to this view, answering by \quotes{filling the blank} of a pre-defined template answer (as most of One-Size-Fits-All approaches do) prevents the act of answering from being explanatory, by lacking \textit{illocution}. These conclusions are quite clear and explicit in Achinstein's last works \cite{achinstein2010evidence}, consolidated after a few decades of public debates.
	
	More precisely, according to Achinstein's theory, an explanation can be summarized as a pragmatically correct content-giving answer to questions of various kinds, not necessarily linked to causality.
	In some contexts, highlighting logical relationships may be the key to making the person understand. In other contexts, pointing at causal connections may do the job. And in still further contexts, still other things may be called for.
	
	As consequence we can see a deliberate absence of a taxonomy of questions (helpful to categorize and better understand the nature of human explanations) to refer.
	This apparently results in a refusal to define a quantitative way to measure how pertinent an answer is to a question, justified by the important assertion that explanations have a pragmatic character, so that what exactly has to be done to make something understandable to someone may (in the most generic case) depend on the interests and background knowledge of the person seeking understanding \cite{douven2012peter}.
	
	\edit{In this sense, the strong connection of Achinstein's theory to natural language and (natural) users is quite evident, for example in the Achinsteinian concepts of:
	\begin{itemize}
		\item \textbf{Ellipses} or \textbf{elliptical information} (\cite{achinstein2010evidence}, pp. 112-114): intended as an explanation that is purposely shrunk to a very minimal sentence to avoid information that might be redundant for the explainee (i.e. for his/her background knowledge or common-sense).
		\item \textbf{U-restrictions} (\cite{achinstein2010evidence}, pp. 114-119): the meaning of an utterance/explanation u is restricted to the common interpretation of it, usually defined by grammar or rhetoric.
	\end{itemize}
	Indeed, according to Achinstein (\cite{achinstein1983nature}, pp. 48-53) \quotes{S explains Q to E by uttering U} is true if and only if either:
	\begin{itemize}
		\item U is constructed in a way that allows anyone to easily \textit{restrict} (i.e. disambiguate, interpret) the meaning of U to that of a sentence expressing a complete content-giving proposition with respect to Q.
		\item U is elliptical (or an \textit{ellipsis}): it is enough for the specific E to understand a sentence expressing a complete content-giving proposition with respect to Q.
	\end{itemize}
	In other words, Achinstein's definition of explanation takes under consideration not only the typical omissions of content that are possible by virtue of grammar and rhetoric (i.e. co-references, anaphora, etc.) but also all those omissions of information used in order to simplify an explanation, reducing the amount of information that is considered redundant for a specific explainee or common-sense.}
	
	\edit{Despite this deep connection to natural (non-formal) language, Achinstein does not reject at all the utility of formalisms, hence suggesting the importance of following \textit{instructions} (protocols, rules, algorithms) for correctly explaining some specific things within specific contexts.
	In this sense, Achinstein's concept of instructions, in particular (\cite{achinstein1983nature}, pp. 53-56), could be usefully adopted to address the question of how deep or broad explanations should go.
	In fact, instructions are \quotes{rules imposing conditions on answers to a question}, or also a mechanism to check whether an answer is correct\footnote{Please note that Achinstein stresses the fact that a correct answer does not necessarily produce understanding. So, correctness is not a sufficient condition for an answer to be an explanation.} in a given context, and they might be framed with legal requirements (as in the case of \ac{XAI} \cite{bibal2021legal}), ethical guidelines (i.e. \cite{hleg2019ethics}), mathematics, etc..}
	
	\edit{\subsection{Explaining as Question Answering in XAI}}
	Overall, the idea of answering questions as explaining is not new to the field of \ac{XAI} and it is also quite compatible with everyone's intuition of what constitutes an explanation.
	
	Two distinct types of explainability are predominant in the literature of eXplainable AI: rule-based and case-based.
	Rule-based explainability is when the explainable information is a set of formal logical rules describing the inner logic of a model, its chain of causes and effects, how it behaves, why that output given the input, what would happen if the input were different, etc.
	While case-based explainability is when the explainable information is a set of input-output examples (or counter-examples) meant to give an intuition of the model's behaviour, i.e. counterfactuals, contrastive explanations, or prototypes\footnote{Prototypes are instances of the ground-truth considered to be similar to a specific input-output for which the similarity explains the model's behaviour.}, etc..
	
	Despite the different types of explainability one can choose, it appears to be always possible to frame the information provided by explainability with one or (sometimes) more questions.
	In fact, it is common to many works in the field \cite{ribera2019can,lim2009and,miller2018explanation,gilpin2018explaining,dhurandhar2018explanations, wachter2017counterfactual,rebanal2021xalgo,jansen2016s,madumal2019grounded} the use of archetypal (e.g. why, who, how, when, etc.) or more specific questions to clearly define and describe the characteristics on explainability, regardless its type.
	
	For example, \citeauthor{lundberg2020local}\cite{lundberg2020local} assert that the local explanations produced by their TreeSHAP (an \textit{additive feature attribution} method for feature importance) may enable \quotes{agents to predict \textit{why} the customer they are calling is likely to leave} or \quotes{help human experts understand \textit{why} the model made a specific recommendation for high-risk decisions}.
	
	While \citeauthor{dhurandhar2018explanations}\cite{dhurandhar2018explanations} clearly state that they designed CEM (a method for the generation of counterfactuals and other contrastive explanations) to answer the question \quotes{why is input x classified in class y?}.
	
	Also, \citeauthor{rebanal2021xalgo}\cite{rebanal2021xalgo} propose and studies an interactive approach where explaining is defined in terms of answering why-what-how questions.
	
	These are just some examples, among many, of how Achinstein's theory of explanations is already implicit in existing \ac{XAI} literature, highlighting how deep is in this field the connection between answering questions and explaining.
	A connection that has been implicitly identified also by \cite{lim2009and}, \cite{miller2018explanation} and \cite{gilpin2018explaining} that analysing \ac{XAI} literature were able to hypothesise that a good explanation, about an automated decision-maker, answers at least the following questions: 
	\begin{itemize}
		\item What did the system do?, 
		\item Why did the system do P?, 
		\item Why did the system not do X?, 
		\item What would the system do if Y happens? , 
		\item How can I get the system to do Z, given the current context?
		\item What information does the system contain?
	\end{itemize}
	
	Nonetheless, despite its compatibility, practically none of the works in \ac{XAI} ever explicitly mentioned Ordinary Language Philosophy's theories, preferring to refer Cognitive Science's \cite{miller2018explanation,hoffman2018metrics} instead.
	This is probably because Achinstein’s \textit{illocutionary} theory of explanations is seemingly difficult to be implemented into a software, by being utterly pragmatic and by missing a precise definition of \textit{illocution} as intended for a computer program.
	In fact, \emph{user-centrality} is challenging and sometimes not clearly connected to \ac{XAI}'s main goal of \quotes{opening the black-box} (e.g. understanding how and why an opaque AI model works).
	
	User-centrality (or pragmatism) in explanations imply the generation of explanatory content specifically tailored to fit the explainee's needs and goals.
	As consequence, considering the latent unpredictability of any generic human explainee, achieving user-centrality is a daunting task requiring a proper understanding of the recipient of the explanation amid constantly mutating background knowledge and objectives.
	Given the complexity of generating user-centred explanations, it is common in Computer Science literature, and especially in \ac{XAI}, to encounter non-user-centred, one-size-fits-all, explanatory tools instead. 
	
	Key to the one-size-fits-all approach is to choose in advance what to tell in an explanation, regardless the needs of the users, by answering well to just one (or sometimes few) pre-defined questions.
	This may become a problem when \ac{XAI}-generated explanations alone have to be deployed in real-world applications, to real end users (i.e. lay persons, or domain experts such as doctors, bankers, judges, drivers).
	
	In fact, compared to creating explanations for AI experts, generating explanations for end users is more challenging, since it is unrealistic to ask all the end users to interpret the internal parameters and complex computations of AI models, having also a diverse range of needs and requirements of using \textit{XAI} systems \cite{jin2019bridging}.
	For example\footnote{We point the reader to the sketches presented in \cite{jin2019bridging} for more examples of how end users may have complex needs to satisfy.}, a lay person trying to receive a loan might be definitely interested in knowing that her/his application was rejected (by an AI) mainly because of an elevated number of inquiries on her/his accounts (as both TreeShap and CEM can say), but this information alone may not be enough for her/him to reach her/his goals. These goals may be out of the scope of the \ac{XAI}, as to understand: how to effectively reduce the number of inquiries in order to get the loan, which types of inquiries may affect his/her status (the hard or the soft ones?), etc.. 
	
	A valid attempt to understand what constitutes a pragmatic explanatory process in the context of \ac{XAI} is probably given by \citeauthor{madumal2019grounded}\cite{madumal2019grounded}.
	To this end, \citeauthor{madumal2019grounded} formalize a model of explanatory process using an \textit{agent dialogue framework}, analysing a few hundreds human-human and human-agent interactions, through the lens of \textit{grounded theory}.
	
	Not surprisingly, considering the adopted ordinary-language-oriented approach, the final model framed by \cite{madumal2019grounded} consists in an iterative question answering process involving also argumentation, but not capturing \textit{illocution}, furthermore without discussing the practical implementation of an algorithm, and considering only a small range of possible explanatory contents focused on causes, justifications and processes.
	We believe that this last bias is probably due to the intrinsic nature of the considered human-human interactions, that were partly Reddit \quotes{gossip} chats (more about frivolous and \textit{non-illocutionary} question answering than explaining) and partly very technical explanatory dialogues (supreme court transcripts, journalistic interviews on politics, finance and computer science) mainly pursuing teleological and causal explanations.
	
	Interestingly, (and indirectly) on the same line of \cite{madumal2019grounded}, also \citeauthor{rebanal2021xalgo}\cite{rebanal2021xalgo} propose and study (only through a Wizard-of-Oz test though) an interactive approach using question answering, to explain deterministic algorithms to non-expert users.
	Nonetheless, similarly to \cite{madumal2019grounded}, also \cite{rebanal2021xalgo} focus on a small sub-set of possible types of explanations, avoiding \textit{illocution}, as suggested by a few of the comments given by their participants: \quotes{it answers everything accurately and it gives the information that I asked for but it does so like sounding more like a glossary like a dictionary}, \quotes{... like a robot’s answers ... If I asked someone to explain it, it wouldn’t give me all this}.
	

	
	\section{Proposed Solution: User-Centred Explanations via Illocutionary Question Answering} \label{sec:proposed_solution}
	Pragmatically explaining to humans is a challenging task, especially for a machine.
	Just to give an intuition, being able to construct useful explanations is one of the main challenges of making science.
	
	The point is that explaining is not just correctly answering a given question but it is also answering all the other implicit questions defined by, e.g., the background knowledge of the explainee, the objectives of the explanatory process, and the given context. It is, in some sense, attempting to anticipate the (conceivably mostly unknown) needs for an explanation by providing, as an archetypal answer, (possibly expandable) summaries of pertinent information. 
	
	Our own approach is based on Achinstein's theory of explanations (1983) \cite{achinstein2010evidence}, where explanations are the result of an \textit{illocutionary} act of pragmatically answering to a question. 
	But what is \textit{illocution} and how is pragmatism achieved within an explanatory process?
	
	Considering that pragmatism is intended as a synonym for \textit{user-centrality}, it can be achieved within an explanatory process through a sufficient amount of usability.
	In short, we adopt the definition of usability as the combination of \emph{effectiveness}, \emph{efficiency}, and \emph{satisfaction}, as per ISO 9241-210, that defines usability as the \quotes{extent to which a system, product or service can be used by specified users to achieve specified goals with effectiveness, efficiency and satisfaction in a specified context of use} \cite{international2010ergonomics}. 
	
	\edit{Effectiveness (\quotes{accuracy and completeness with which users achieve specified goals}) and efficiency (\quotes{resources used in relation to the results achieved. [\dots] Typical resources include time, human effort, costs and materials.}) can be assessed through objective measures (in our case, pass vs. fail at domain-specific questions and time to complete tasks, respectively). 
	Satisfaction, defined as \quotes{the extent to which the user's physical, cognitive and emotional responses that result from the use of a system, product or service meet the user’s needs and expectations}, is a subjective component and it needs a direct confrontation with the user.
	Satisfaction is normally measured with standardised questionnaires. 
	One of these is \acf{SUS} \cite{brooke2013sus}, that (despite its sometimes confusing name) is used to measure the subjective satisfaction (or perceived usability) and not the usability (that according to the ISO standard is the combination of both objective and subjective metrics: effectiveness, efficiency and satisfaction) \cite{borsci2015assessing}.
	Importantly, \ac{SUS} is considered one of the most widely used standardized questionnaire for the assessment of post-test satisfaction \cite{sauro2009correlations,albert2013measuring,lewis2018measuring}.}
	
	What is of utmost importance for a proper user-centrality is to help the user in the process of achieving her/his own goals.
	If we agree on Achinstein's interpretation of explanations, in an explanatory process the goals of a user are identified by questions.
	Some questions may be explicit and others not, some may lose importance over time or vice-versa, but normally a user is fully satisfied with explanations only when they efficiently convey a full coverage of pertinent answers for all his/her objectives.
	
	Hence, considering that pragmatism is achieved when explanations meet the user's goal, any good explanatory tool should provide reasonable mechanisms for the explainee to specify his/her own questions. 
	Problems arise when these questions are not explicitly posed, requiring the explanatory tool to infer them automatically.
	In fact, it is certainly not trivial to correctly elicit the user's implicit goals, and sometimes it is not even easy for the user to express or understand goals in an intelligible or precise way.
	
	We assert that, in human-generated explanations, \textit{illocution} is the main mechanism responsible for anticipating unposed questions, shaping the underlying explanatory process as more user-centred and helping both the explainee and the explainer in consuming less resources, reducing the amount of explanatory steps.
	\edit{Indeed, we believe that, in the most generic case, \textit{illocution} in explanations is equivalent to the act of \textit{pertinently and deliberately answering} to implicit questions characterised by the user, and that is different from the Achinsteinian concept of \textit{instructions} but in some way akin to the Achinsteinian concept of \textit{ellipsis} briefly introduced in Section \ref{sec:background}.} 
	
	\edit{\begin{definition}[Illocution in Explaining] \label{def:illocution}
		Explaining is an illocutionary act which involves providing answers to an explicit question on some topic together with several other implicit, or unposed, questions that are deemed to be necessary for the explainee to properly understand the topic.
		Sometimes these implicit questions can be inferred through a thorough analysis of the explainee's background knowledge, history and objectives, considering also Frequently Asked Questions (FAQs).
		But, in the most generic case, when no assumption can be done on the explainee's knowledge and objectives, the only implicit questions that is possible to exploit for \textit{illocution} are the most generic ones, called \textit{archetypal questions}.
	\end{definition}}
	
	\edit{\begin{definition}[Archetypal Question] \label{def:archetypal_question}
		An archetypal question is an archetype applied on a specific aspect of something to be explained (or explanandum, in Latin). 
		Examples of archetypes are the interrogative particles (why, how, what, who, when, where, etc.), or their derivatives (why-not, what-for, what-if, how-much, etc.), or also more complex interrogative formulas (what-reason, what-cause, what-effect, etc.).
		Accordingly, the same archetypal question may be rewritten in several different ways, as \quotes{why} can be rewritten in \quotes{what is the reason} or \quotes{what is the cause}.
		In other terms, archetypal questions identify generic explanations about a specific aspect to explain (e.g. a topic, an argument, a concept,etc.), in a given informative context. 
	\end{definition}}

	For example, if the explanandum would be \quotes{heart diseases}, there would be many aspects involved including \quotes{heart}, \quotes{stroke}, \quotes{vessel}, \quotes{diseases}, \quotes{angina}, \quotes{symptoms}, etc..
	Some archetypal questions in this case might be \quotes{What is an angina?} or \quotes{Why a stroke?}, etc..
	
	More specifically, in an explanatory process about a fixed explanandum, when a precise initial question is provided by the explainee, \textit{illocution} is embedded in the consequent explanation through digressions, answering other implicit questions (i.e. the archetypal ones). 
	On the other side, when no question is given by the explainee but an explanandum, \textit{illocution} is about providing an overview as aggregation of different answers to implicit questions about the aspects of that explanandum, as in the example of the \quotes{university lectures} described in Section \ref{sec:introduction}. 
	
	The archetypal questions prevent by design any \quotes{filling the blank} answer, thus meeting the tricky but reasonable assumption of \textit{illocution} given by Achinstein for his pragmatic theory of explanations. 
	\edit{\textit{Illocution} is, in some sense, attempting to anticipate the (conceivably mostly unknown) explainee's needs for an explanation by providing, as an archetypal answer, possibly expandable summaries of (more detailed) pertinent information. 
	In other terms, we believe that the more explicit and implicit questions are answered by an explanatory process, the more likely the resulting explanations are going to meet the explainee's objectives, the more usable (effective, efficient and satisfactory) the explanatory tools.}
	
	\edit{Therefore we have the following hypothesis.
	\begin{hypothesis}[Main] \label{hyp:main}
		If the following premises are true:
		\begin{itemize}
			\item an explanatory process is an illocutionary act of providing content-giving answers to questions;
			\item illocution is about correctly answering not just to some explicit question but also to all the implicit questions that the explainee might need.
		\end{itemize}
		Then, given an arbitrary explanatory process, increasing its \textit{goal-orientedness} and/or \textit{illocutionary power} results in the generation of more usable explanations.
		Where the \textit{goal-orientedness} of an explanatory process is its ability to answer the \textit{explicit} questions of an explainee, and the \textit{illocutionary power} is its ability to anticipate and answer the \textit{implicit} (archetypal) questions of an explainee.
	\end{hypothesis}}

	
	To verify this hypothesis we designed a proof of concept algorithm for illocutionary question answering and a couple of experiments on different explananda.
	
	\section{Proof of Concept: An Algorithm for Generating Explanations} \label{sec:proof_of_concept}
	\edit{From the definition of \textit{illocution} given in Section \ref{sec:proposed_solution}, it follows that illocutionary question answering requires a mechanism for pragmatically:}
	\begin{enumerate}
		\item estimating the pertinence of answers to (archetypal) questions,
		\item identifying the set of relevant aspects to be explained through \textit{illocution}.
	\end{enumerate}
	The problem with this is that every user may need different informative contents depending on her/his background knowledge, therefore making very hard to estimate what is the pertinence of an informative content, at least pragmatically speaking.
	
	To solve this problem, we frame \textit{pertinently answering} as the process of giving (archetypal) answers that are likely to be pertinent for a given (archetypal) question. The likelihood can be quantitatively estimated on strong-enough statistical evidence collected from large corpora and built in language models. 
	The point is that, this statistical definition of pertinence 
	is compatible with Achinstein's \textit{u-restrictions} (introduced in Section \ref{sec:achinstein}) and it 
	does not preclude a pragmatic (user-centred) explanatory process that is locally non-pragmatic but globally pragmatic.
	
	In fact, we might see the space of all the explanations about an explanandum (or \acl{ES} \cite{sovrano2020modelling}) as a sort of manifold space where every point within it is interconnected explainable information that is not user-centred locally (because it is the same for every user), but globally as an element of a sequence of information that can be chosen by users according to their interest drifts while exploring the space.
	
	\edit{Importantly, this understanding of an explanation as a sequence within the \ac{ES} is indeed framing explanations as \textit{ellipses} (a concept introduced in Section \ref{sec:achinstein}), for the explanation being a pragmatic subset of all the possible information about the explanandum.}
	
	Our proof of concept builds over the extraction and structuration of an \acf{ES} \cite{sovrano2020modelling}, intended as all the possible explanations (about an explanandum) reachable by a user: 
	\begin{itemize}
		\item through an explanatory process, 
		\item starting from an initial explanation, 
		\item via a pre-defined set of actions. 
	\end{itemize}
	According to the model of \citeauthor{sovrano2020modelling}, we might see the \ac{ES} as a graph of interconnected bits of explanation, and an explanation as nothing more than a path within the \ac{ES}. 
	
	Consequently, the relevant aspects to be explained are framed as clusters of these interconnected bits of explanation.
	Assuming that the explanandum is a set of documents written in a natural language (e.g. English), the relevant aspects to explain might be (for example) the different concepts/entities within the corpus, so that to each concept it is possible to associate an overview; e.g. in the sentence \quotes{the customer opened a new bank account} different entities are \quotes{customer}, \quotes{bank}, \quotes{bank account}.
	
	The choice of an initial explanation is generally dependent on the nature of the explanandum and the objectives associated with the category of users involved in the explanatory process. A good choice of initial explanation could be an overview of the whole explanandum or of the explanatory process. Therefore, in the case of \ac{XAI}, a proper initial explanation might be the static explanation provided by the \ac{XAI} algorithm (e.g. by compiling a template or generating text through a formal language).
	
	In order for a user to explore such \ac{ES} through an explanatory process, a pre-defined set of interactions has to be identified.
	As primitive actions, according to our understanding of Achinstein's theory, we might consider: 
	\begin{itemize}
		\item \textbf{Open Question Answering}: the user writes a question and then it gets one or more relevant punctual answers.
		\item \textbf{Aspect Overviewing}: the user selects an aspect of the explanandum (i.e. contained in an answer) receiving as explanation a set of \textit{relevant} archetypal answers involving other different aspects that can be explored as well. Archetypal answers can also be expanded, increasing the level-of-detail.
	\end{itemize}
	In other terms, we can see any overview or explanatory answer as an appropriate paraphrase of a sequence of \acf{ES} points.
	
	These \aclp{ES} can be very complex graphs, making the exploration a very challenging task for a human. To tackle this issue, one way to go is to decompose the \ac{ES} in a tree\footnote{In graph theory, tree decompositions are used to speed up solving certain computational problems on graphs. Practically speaking, many instances of NP-difficult problems on graphs can be efficiently solved via tree decomposition \cite{bodlaender1997treewidth}.}. 
	To do so, some heuristics shall provide a policy for at least:
	\begin{inparaenum}[i)]
		\item organising the \ac{ES}'s nodes or aspects,
		\item structuring the information internal to the \ac{ES}'s nodes,
		\item ordering/filtering the \ac{ES}'s edges in a way that would effectively decompose the graph into a tree.
	\end{inparaenum}
	
	The heuristics we adopted are inspired by \cite{sovrano2020modelling} and they are respectively:
	\begin{itemize}
		\item \textbf{Abstraction}: for identifying the explanandum's aspects, or \ac{ES}'s nodes,
		\item \textbf{Relevance}: for organising the information internal to the \ac{ES}'s nodes,
		\item \textbf{Simplicity}: for filtering the information internal to \ac{ES}'s nodes and also for selecting the viable \ac{ES}'s edges.
	\end{itemize}
	We will refer to them as the ARS heuristics.
	
	In order to implement the three aforementioned heuristics and the primitive actions, extracting an \ac{ES}, identifying the set of relevant aspects to be explained through \textit{illocution} and estimating the pertinence of information, we may use an algorithm like the one proposed by \cite{sovrano2020legal} for efficient question answer retrieval.
	\begin{figure}[!htb]
		\centering
		\includegraphics[width=.9\columnwidth]{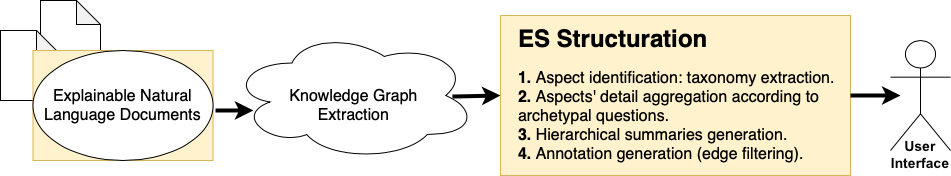}
		\caption{\textbf{The Pipeline}: A simple diagram summarising the pipeline of our user-centric explanatory software.}
		\label{fig:yai_diagram}
	\end{figure}
	As shown in figure \ref{fig:yai_diagram}, this algorithm would:
	\begin{enumerate}
		\item Identify and extract out of the explanandum all the different aspects (concepts/entities) and their related information.
		\item Build a knowledge graph so that concepts/entities are linked together.
		\item Extract a taxonomy from the knowledge graph.
		\item Build one or more information clusters for every aspect, according to the identified archetypal questions.
		\item Order the information within the clusters according to its pertinence to archetypal questions.
		\item Filter the external edges of the \ac{ES}, favouring shorter and simpler paths/explanations, thus reasonably reducing the amount of redundant information for a human.
	\end{enumerate}
	
	\subsection{Knowledge Graph Extraction} \label{sec:kg_extraction}
	\ac{KG} extraction is the extraction of concepts and their relations, from natural language text, in the form of a graph where concepts are nodes and relations are edges. 
	We are looking for a way to extract \acp{KG} that somehow preserve the original natural language, preferring them over classical \ac{RDF} graphs. This way we can easily make them inter-operate with deep-learning based \ac{QA} algorithms and existing language models.
	
	More in detail, as in \cite{sovrano2020legal}, we perform \ac{KG} extraction by:
	\begin{enumerate}
		\item Analysing the grammatical dependencies of the tokens extracted by Spacy's Dependency Parser, thus identifying the (target) concepts and entities in the form of syntagms.
		\item Using the dependency tree to extract all the tokens connecting two different target concepts in a sentence, thus building a textual template formed by the ordered sequence of the identified tokens and the target concepts replaced with the placeholders \quotes{\{subj\}} and \quotes{\{obj\}} (in accordance with their grammatical dependencies).
		\item Creating a graph of subject-predicate-object triples where the target concepts are the subject and the object and the textual template is the predicate.
	\end{enumerate}
	
	The resulting triples are not standard triples\footnote{This is why we are using the method proposed in \cite{sovrano2020legal}}. In fact, they are a sort of function, where the predicate is the body of the function and the object/subject are its parameters. Obtaining a natural language representation of these template-triples is straightforward by design, by replacing the instances of the parameters in the body. An example of such a template-triple (in the form subject, predicate, object) is:
	\begin{inparadesc}
		\item \quotes{the applicable law},
		\item \quotes{Surprisingly \{subj\} is considered to be clearly more related to \{obj\} rather than to something else.},
		\item \quotes{that Member State}.
	\end{inparadesc}
	
	Therefore, to increase the interoperability of the extracted \ac{KG} with external resources we performed the following extra steps:
	\begin{inparaenum}[i)]
		\item We automatically assigned a URI and a RDFS label to every node of the graph. The URI is obtained by lemmatising the label.
		\item We automatically added special triples to keep track of the snippets of text (the sources) from which the concepts and the relations are extracted.
		\item We automatically added sub-class relations between composite concepts (syntagms) and the simpler concepts composing the syntagm.
	\end{inparaenum}
	
	Because of the adopted extraction procedure, the resulting \ac{KG} is not perfect, containing mistakes caused by wrong dependency assignments or similar issues. Despite this, due to the fact that the original natural language is practically preserved thanks to the textual templates, this will not impact significantly on \ac{QA}.
	
	\subsection{Taxonomy Construction: Nodes Clustering} \label{sec:taxonomy_construction}
	In order to efficiently use, query and explore the extracted \ac{KG}, we need to structure it in a proper way.
	We believe that effective abstract querying can be possible by structuring the \ac{KG} as a light ontology, giving it a solid backbone in the form of a taxonomy. 
	In fact, being able to identify the types/classes of a concept would allow to perform queries with a reasonable level of abstraction, making possible to refer to all the sub-types (or to some super-types) of a concept without explicitly mentioning them.
	
	The taxonomy construction phase consists in building one or more taxonomies, via \acf{FCA} \cite{ganter2012formal}.
	In order to build a taxonomy via \ac{FCA} one approach consists in exploiting, as \ac{FCA}'s properties, the hypernyms relations of the concepts in the \ac{KG}. 
	We found that the simplest way to extract such relations is through the alignment of the \ac{KG} to WordNet\footnote{We are aware that WordNet is not omni-comprehensive, but at this stage of the work we are only interested in extracting a reasonable taxonomy.}, through a Word-Sense Disambiguation algorithm.
	
	The application of \ac{FCA} on the aligned WordNet concepts (and their respective hypernyms) produces as result a forest of taxonomies. Every taxonomy in the forest is a cluster of concepts rooted into very abstract WordNet concepts that we can use as label for the respective taxonomies.
	
	\subsection{Overview Generation via Question Answering: Information Clustering and Summarisation} \label{sec:qa}
	As mentioned in the previous sections, we can generate an overview by clustering and ordering information with respect to its pertinence to a set of archetypal questions.
	
	The essential idea is to generate a concept overview by performing \ac{KG}-based question answering, retrieving the most similar concept's triples for each archetype.
	\ac{KG}-based question answering consists in answering natural language questions about information contained in the \ac{KG}. More in detail, let $Q$ be an archetypal question and $C$ a concept, we perform information clustering by:
	\begin{enumerate}
		\item \textbf{Extracting} all the template-triples related to $C$, including those of $C$'s sub-classes.
		\item \textbf{Selecting}, among the natural language representations of both the retrieved triples and their respective subjects/objects, the snippets of natural language that are sufficiently likely to be an answer to $Q$.
		\item \textbf{Returning} as set of answers the contexts (the source paragraphs) of the selected triples, ordered by pertinence.
	\end{enumerate}
	
	More in detail, the \textit{selecting} phase is performed by computing the pertinence of an answer as the inner product between the embeddings of the contextualised snippets of text and the embedding of $Q$. The aforementioned embedding is obtained by means of a specialised language model such as the \acf{USE} for \ac{QA} \cite{yang2019multilingual}, while the context is the source paragraph from which a snippet of text is extracted from the original document. If a snippet of text has a similarity above a given threshold, then it is said to be sufficiently likely an answer to $Q$, therefore pertinent.
	
	Considering that an answer could be reasonably associated to more than one archetypal question, we decided to apply an heuristic filtering strategy in the attempt to minimise redundant information, thus following the simplicity heuristic. 
	To do so, we had to attempt a sort of hierarchical organisation of the archetypes, defining some questions as more generic than others, thus prioritising the less generic ones. 
	
	In fact, in some cases an answer to the question \quotes{What?} could also be a valid answer to \quotes{What for?}. 
	This is because \quotes{What for?} is intuitively more specific than \quotes{What?}. Hence, to reduce redundancy, we can force answers to be exclusive to a single archetypal question, assigning first the answers to the most specific archetypes.
	A descending ordering of specificity, that we found meaningful for the identified archetypal questions, is: 
	\begin{inparadesc}
		\item what, 
		\item why, 
		\item what-for, 
		\item how, 
		\item who, 
		\item where, 
		\item and when.
	\end{inparadesc}
	Such ordering seemed to be proper for the purposes of the proof of concept presented in Section \ref{sec:proof_of_concept}, but it is likely that a different ordering is required for different purposes.
	
	Finally, after the identification of a set of answers for a question $Q$, we can build an expandable summary by recursively concatenating together few answers and by summarising them (thus recursively building a tree of summaries) through one of the state-of-the-art deep learning algorithms for extractive or abstractive summarisation provided by \citeauthor{wolf2019transformers}\cite{wolf2019transformers}.
	At the end of the process we have that an overview is defined by a (sometimes empty) expandable summary for every archetypal question, plus the list of super-classes, sub-classes, sub-types (if any) and eventually few other external resources considered to be of any use (e.g. a short abstract of few words).
	
	\edit{Therefore, we have that the additional taxonomical information is used for the abstraction policy, while the rest of the information is meant to be used for both the relevance and the simplicity policies.}
	
	\subsection{Overview Annotation: Edge Filtering} \label{sec:overview_annotation}
	Every sentence in the overview is annotated. Annotations consist in linking a concept's embodiment to its corresponding overview (so that clicking on the link would open the overview). 
	
	The edge filtering algorithm has to decide which syntagms to annotate, in order to avoid annotating every possible concept expressed in a sentence, including redundant or useless ones. More precisely, the edge filtering algorithm would remove:
	\begin{itemize}
		\item Those concepts that can be assumed of scarce relevance for a common user, as those likely to be already known by someone with a basic understanding of English (examples are: day, time, space, November, etc..). These concepts are associable to generic world-knowledge and they can be heuristically identified by analysing the words frequency in the Brown corpus \cite{francis1979brown} or similar corpora.
		\item The concepts with a betweenness centrality equal to $0$. In fact, filtering these concepts would reduce the average length of an explanation (intended as a path over the \ac{ES}) without preventing the user from reaching the information it needs.
	\end{itemize}
	
	\section{Experiment} \label{sec:experiments}
	Regardless of the tool for explaining that we adopt (i.e. the one we described in section \ref{sec:proof_of_concept}), or the direction we take to produce explanation, we aim to prove that the usability (as per ISO 9241-210) of an explanatory process can be affected by its \textit{illocutionary power} and \textit{goal-orientedness}\footnote{See Hypothesis \ref{hyp:main} for a definition of both \textit{illocutionary power} and \textit{goal-orientedness}.}.
	
	\edit{In order to verify hypothesis \ref{hyp:main}, we test our algorithm on two different explananda consisting in \ac{XAI}-powered systems (for credit approval and heart disease prediction) and probe into it from the perspective of different users. 
	More in detail, we compare three different explanatory approaches to present such systems to the users:
	\begin{itemize}
		\item \textbf{Overwhelming Static Explainer (OSE, in short)}: a fully static one-size-fits-all explanatory tool that does not attempt to answer any implicit or explicit question, dumping on the user large portions of text that could only possibly contain the information sought.
		\item \textbf{How-Why Narrator (HWN, in short)}: an interactive version of OSE designed to provide causal and expository explanations, answering exclusively to \quotes{how} and \quotes{why} archetypal questions, not allowing the users to ask their own. Therefore HWN has no \textit{goal-orientedness} and little \textit{illocutionary power}. 
		\item \textbf{Explanatory AI for Humans (YAI4Hu, in short)}: a more interactive version of HWN. YAI4Hu is designed to have much greater \textit{goal-orientedness}, empowering the users with the ability to ask their own questions through \textit{Open Question Answering}. Furthermore, YAI4Hu has also more \textit{illocutionary power}, answering not just to \quotes{how} and \quotes{why} archetypal questions but also to \quotes{what} questions and many others.
	\end{itemize}
	In short, the difference between HWN and YAI4Hu is the amount of \textit{illocutionary power} and \textit{goal-orientedness} involved. This should help verifying our hypothesis.
	In fact, the aforementioned interactive explanatory tools are specifically designed to be an extension of their static version (OSE), so that comparing the usability of those tools would indirectly allow us to isolate and measure the effects of \textit{illocution} and \textit{goal-orientedness}.}
	
	\subsection{Explananda} \label{sec:explananda}
	The two explananda are:
	\begin{itemize}
		\item A \textbf{heart disease predictor} based on XGBoost\cite{chen2016xgboost} and TreeShap\cite{lundberg2020local}.
		\item A \textbf{credit approval system} based on a simple Artificial Neural Network and on CEM\cite{dhurandhar2018explanations}.
	\end{itemize}
	
	The credit approval system was designed by IBM to showcase its \ac{XAI} library: AIX360.
	This explanandum is about finance and the system is used by a bank. This bank deploys an Artificial Neural Network to decide whether to approve a loan request, and it uses the CEM algorithm to create post-hoc contrastive explanatory information. This information is meant to help the customers, showing them what minimal set of factors is to be manipulated for changing the outcome of the system from denial to approval (or vice-versa).
	
	The Artificial Neural Network was trained on the \quotes{FICO HELOC} dataset\cite{holterfico}. The FICO HELOC dataset contains anonymized information about Home Equity Line Of Credit (HELOC) applications made by real homeowners. A HELOC is a line of credit typically offered by a US bank as a percentage of home equity. The customers in this dataset have requested a credit line in the range of USD 5,000 - 150,000.
	
	Given the specific characteristics of this system, it is possible to assume that the main goal of its users is about understanding what are the causes behind a loan rejection and what to do to get the loan accepted. 
	The mere output of the \ac{XAI} can answer to the question: \quotes{What are the minimal actions to perform in order to change the outcome of the credit approval system?}. 
	\edit{Nonetheless many other relevant questions might be to answer before the user is satisfied, reaching its goals. 
	Generally speaking, all these questions can be shaped by contextually implicit instructions (for more details see Section \ref{sec:achinstein}) set by specific legal or functional requirements \cite{bibal2021legal}.
	These questions include: \quotes{How to perform those minimal actions?}, \quotes{Why are these actions so important?}, etc..}
	
	On the other hand, the heart disease predictor is a completely new explanandum we designed specifically for the purposes of this paper.
	This explanandum is about health and the system is used by a first level responder of a help-desk for heart disease prevention. 
	The systems uses XGBoost\cite{chen2016xgboost} to predict the likelihood of a patient having a heart disease given its demographics (gender and age), health (diastolic blood pressure, maximum heart rate, serum cholesterol, presence of chest-pain, etc.) and the electrocardiographic (ECG) results. This likelihood is classified into 3 different risk areas: low (probability of heart disease below 0.25), medium ($0.25 < p < 0.75$) or high.
	
	The dataset used to train XGBoost is the \quotes{UCI Heart Disease Data}\cite{detrano1989international,alizadehsani2019database}.
	TreeSHAP\cite{lundberg2020local}, a famous \ac{XAI} algorithm specialised on tree ensemble models (i.e. XGBoost) for post-hoc explanations is used to understand what is the contribution of each feature to the output of the model (XGBoost). 
	TreeSHAP can be used to answer the following questions: \quotes{What are the most important factors leading that patient to this probability of heart disease?}, \quotes{How important is a factor for that prediction?}. 
	
	The first level responder is responsible for handling the patient's requests for assistance, forwarding them to the right physician in the eventuality of a reasonable risk of heart disease.
	First level responders get basic questions from callers, they are not doctors but they have to decide on the fly whether the caller should speak to a real doctor or not. So they quickly use the \ac{XAI} system to figure out what to answer to the callers and what are the next actions to suggest. This system is used directly by the responder, and indirectly by the caller through the responder. 
	These two types of users have different but overlapping goals and objectives. It is reasonable to assume that the goal of the responders is to answer in the most efficient and effective way the questions of the callers.
	To this end, the questions answered by TreeSHAP are quite useful, but many other important questions should also be answered, including: \quotes{What is the easiest thing that the patient could actually do to change his heart disease risk from medium to low?}, \quotes{How could the patient avoid raising one of the factors, preventing his heart disease risk to raise?}, etc..
	
	\subsection{Explanatory Approaches} \label{sec:explanatory_tools}
	The explanatory approaches are:
	\begin{enumerate}
		\item \textbf{OSE}: showing the output of the \ac{XAI} and the whole explanandum exhaustively.
		\item \textbf{HWN}: showing only how-why explanations through \textit{Overviewing}.
		\item \textbf{YAI4Hu}: showing a wide range of archetypal answers (not just how-why ones) through \textit{Overviewing} and allowing \textit{Open Question Answering}.
	\end{enumerate}
	
	The first system is a \textit{2nd-Level Exhaustive Explanatory Closure} or \textit{Overwhelming Static Explainer}  (OSE, in short), a One-Size-Fits-All explanatory tool consisting of two levels of information.
	
	\begin{figure}[!htb]
		\centering
		\includegraphics[width=1\columnwidth]{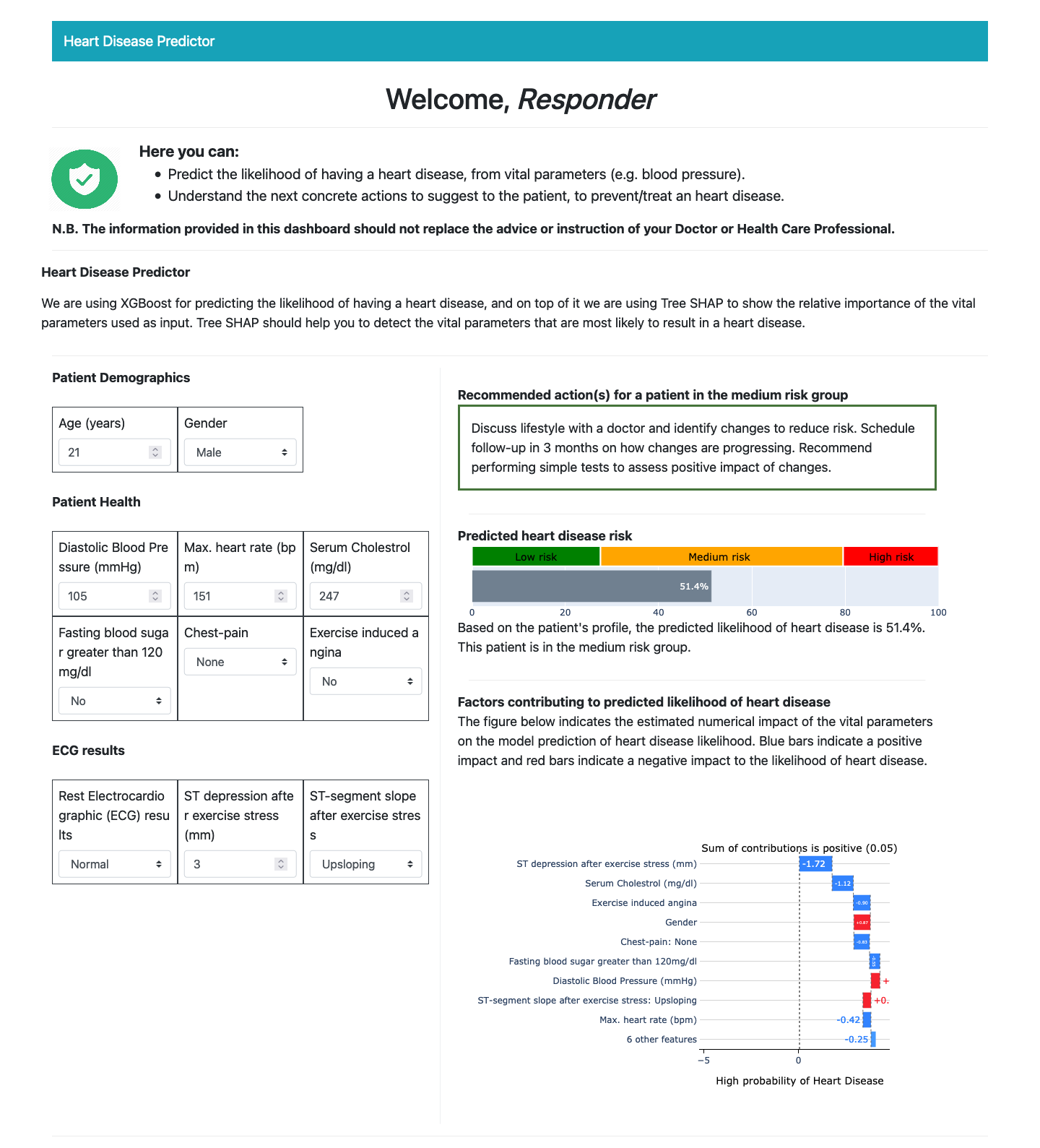}
		\caption{\textbf{Heart Disease Predictor \& OSE}: A screenshot of the OSE explanatory tool for the heart disease predictor.}
		\label{fig:xxx_hd}
	\end{figure}
	The first level (figure \ref{fig:xxx_hd} shows an example for the Heart Disease Predictor) is the initial explanation, providing the bare output of the \ac{XAI} as fixed explanation for all users, together with the output of the wrapped \ac{AI}, extra information to ensure the readability of the results, and a few hyper-links to the second level. 
	
	The second level consists in an exhaustive and verbose set of autonomous static explanatory resources, for the user to understand the explanandum. 
	The information presented at this 2nd level is the content of several resources (e.g. a few hundred web-pages) carefully selected to cover as much as possible of the explanandum topics.
	The OSE is organized therefore as a very long text document (more than 50 pages per system, when printed), with no pragmatic re-organization, besides an automatically created table of content allowing the user to move from the 1st explanatory level to the 2nd.
	
\begin{figure}[!htb]
		\centering
		\includegraphics[width=1\columnwidth]{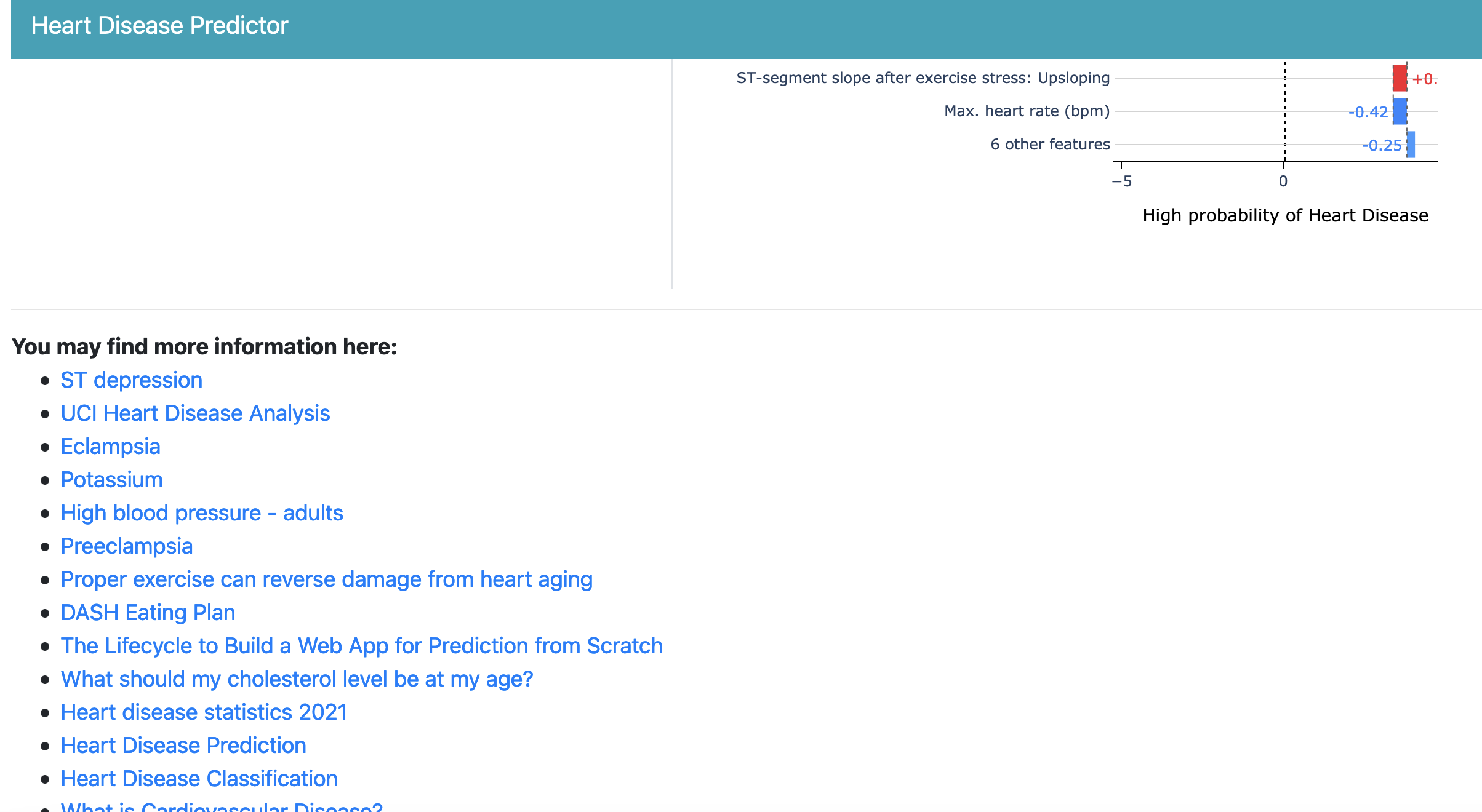}
		\caption{\textbf{Heart Disease Predictor \& OSE}: A screenshot showing the connection between the 1st and the 2nd explanatory levels of OSE on the heart disease predictor.}
		\label{fig:2ec_connection_hd}
	\end{figure}

	The connection between this 2nd level and the 1st level is simply a list of hyper-links to the autonomous resources, appended to the 1st level, as shown in figure \ref{fig:2ec_connection_hd}.
	
	In the case of the heart disease predictor, the first level of OSE consists of:
	\begin{itemize}
		\item \textbf{Context}: a titled heading section kindly introducing the responder (the user) to the system.
		\item \textbf{\ac{AI} Inputs}: a panel for inserting the patient's parameters.
		\item \textbf{\ac{AI} Outputs}: a section displaying the likelihood of heart disease estimated by XGBoost and a few generic suggestions about the next actions for the patient to take.
		\item \textbf{\ac{XAI} Outputs}: a section showing the contribution (positive or negative) of each parameter to the likelihood of heart disease, generated by TreeSHAP.
	\end{itemize}
	While for the second level we take 103 web-pages, 75 of which come from the website of the U.S. Centers for Disease Control and Prevention\footnote{\url{https://www.cdc.gov}}, while the remaining come from the American Heart Association\footnote{\url{https://www.heart.org}}, Wikipedia, MedlinePlus\footnote{\url{https://medlineplus.gov}}, MedicalNewsToday\footnote{\url{https://www.medicalnewstoday.com}} and other minor sources.
	
	A screenshot of the 1st level of OSE for the heart disease predictor is shown in figure \ref{fig:xxx_hd}.
	
	In the case of the credit approval system, OSE consists of\footnote{\edit{The credit approval system does not have the \ac{AI} Inputs because the inputs of the AI are set by the Bank and the user cannot change them from the interface.}}:
	\begin{itemize}
		\item \textbf{Context}: a titled heading section kindly introducing Mary (the user) to the system.
		\item \textbf{\ac{AI} Output}: the decision of the Artificial Neural Network for the loan application. This decision normally can be \quotes{denied} or \quotes{accepted}. For Mary it is: \quotes{denied}.
		\item \textbf{\ac{XAI} Output}: a section showing the output of CEM. This output consists in a minimal ordered list of factors that are the most important to change for the outcome of the \ac{AI} to switch.
	\end{itemize}
	While for the second level we take 58 web-pages, 50 of which come from MyFICO\footnote{\url{https://www.myfico.com}} (the main resource about FICO scores), while the remaining come from Forbes\footnote{\url{https://www.forbes.com}}, Wikipedia, AIX360\footnote{\url{http://aix360.mybluemix.net}}, and BankRate\footnote{\url{https://www.bankrate.com}}.
	
\begin{figure}[!htb]
		\centering
		\includegraphics[width=1\columnwidth]{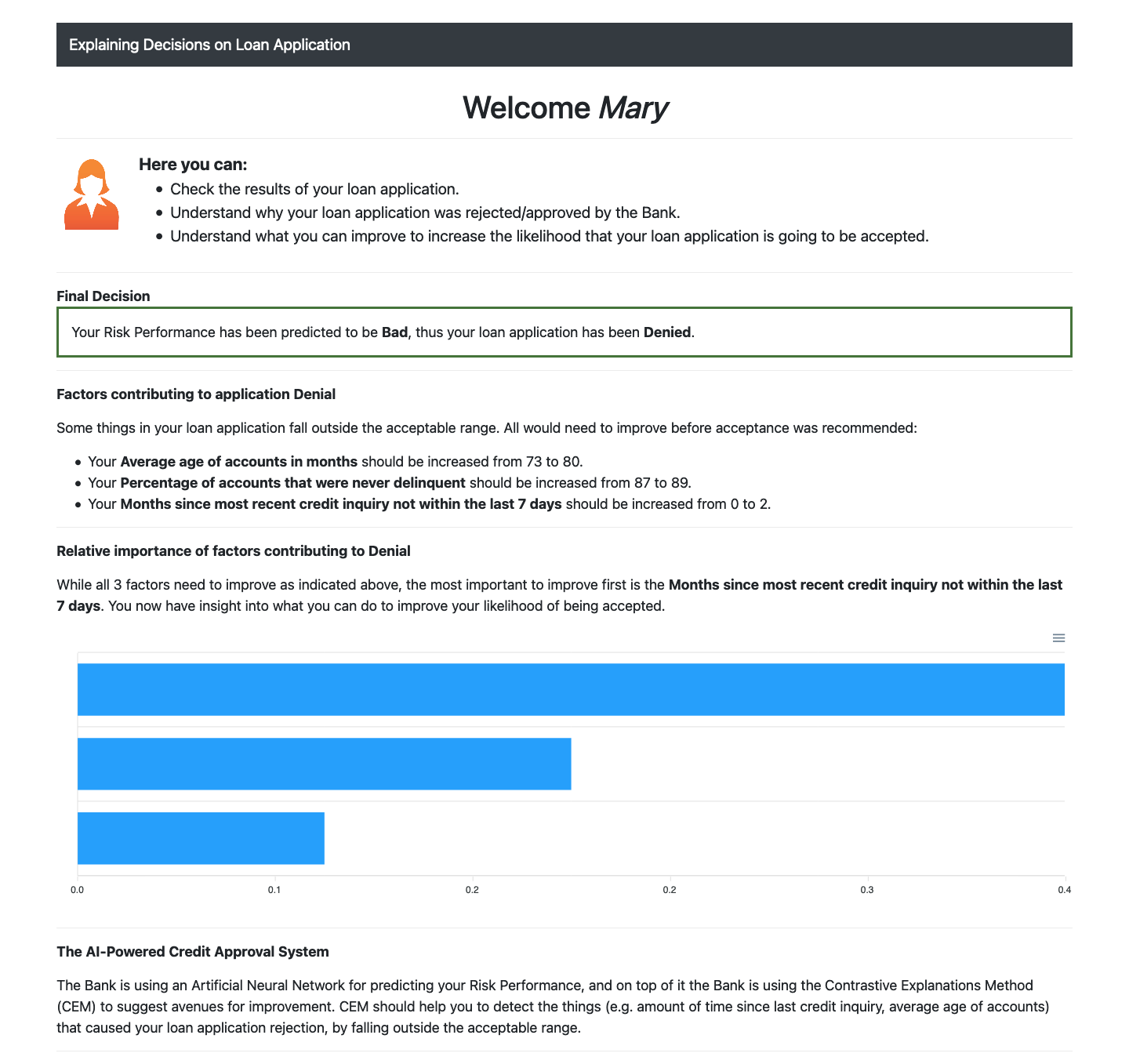}
		\caption{\textbf{Credit Approval System \& OSE}: A screenshot of the OSE explanatory tool for the credit approval system.}
		\label{fig:xxx_ca}
	\end{figure}

	A screenshot of OSE for the credit approval system is shown in figure \ref{fig:xxx_ca}.
	
	We take far more information (almost the double) for the heart disease predictor because, intuitively, it is a more complex explanandum than the credit approval system, requiring much more questions to be covered with different levels of detail.
	
	YAI4Hu is the algorithm described in section \ref{sec:proof_of_concept}. It implements both \textit{Open Question Answering} and \textit{Aspect Overviewing}.
	\textit{Open Question Answering} is for the user to specify its own goals, and it is supposed to be used by those knowing what and how to ask. In other terms, \textit{Open Question Answering} is clearly intended as a mechanism for \textit{locating} information.
	\textit{Open Question Answering} is possible by writing any question in a simple text input at the beginning of the application, connected to a python server exposing the necessary APIs to interact with the pipeline described in \cite{sovrano2020legal}.
	
	On the other hand, \textit{Aspect Overviewing} is a mechanism for \textit{exploring} information and articulating understandings. Through \textit{Aspect Overviewing} a user can navigate the whole \acf{ES} reaching explanations for every identified aspect of the explanandum. 
	In fact, every sentence presented to the user is annotated through a javascript module that makes the text interactive, so that users can select which aspect to overview by clicking on the annotated syntagms.
	\begin{figure}[!htb]
		\centering
		\includegraphics[width=.7\columnwidth]{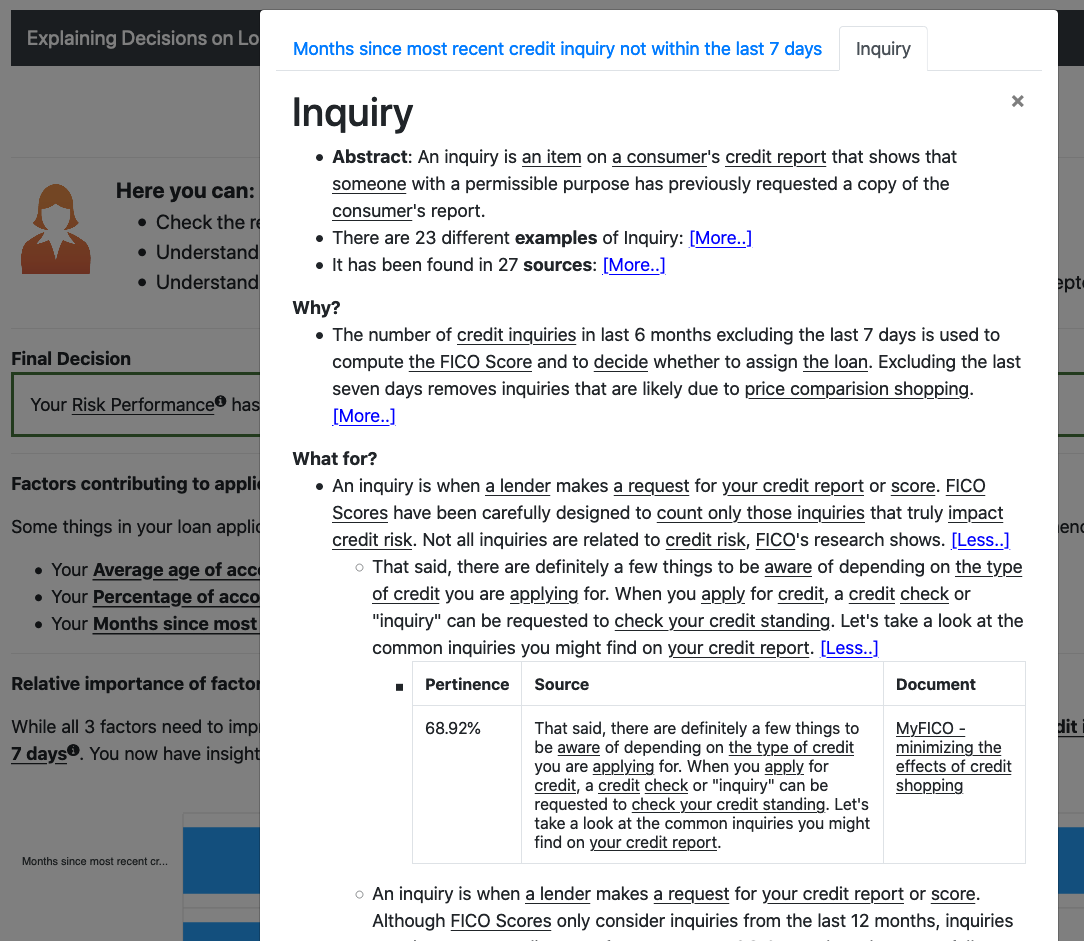}
		\caption{\textbf{Credit Approval System \& HWN}: Example of overview displaying relevant information about a concept that is directly involved in the initial explanation.}
		\label{fig:overview_modal_hd}
	\end{figure}
	
	Annotated syntagms are clearly visible because they have a unique style that makes them easy to recognize, as shown in figure \ref{fig:overview_modal_hd}.
	After clicking on an annotation, a modal opens, showing a card with the most relevant information about the aspect. 
	The most relevant information shown in a card is:
	\begin{enumerate}
		\item A short description of the aspect (if available): abstract and type.
		\item The list of aspects taxonomically connected.
		\item A list of archetypal questions and their respective answers ordered by estimated pertinence. Each piece of answer consists in an \textit{information unit}.
	\end{enumerate}
	
	All the information shown inside the modal is annotated as well. This means (for example) that clicking on the taxonomical type of the aspect, the user can open a new card (in a new tab) displaying relevant information about the type.
	
	The content of the overview modal is obtained by the system by interrogating a python server exposing the necessary APIs to interact with the pipeline described in Section \ref{sec:proposed_solution}.
	The overall extension is designed to be as generic as possible. In other terms it would be possible to use it on any explanatory system providing textual explanations and rich enough documentation (as the Credit Approval System of IBM), because the aforementioned annotation process is fully automated, as described in Section \ref{sec:proposed_solution}.
	
	Differently from OSE, YAI4Hu tries to achieve an Nth-Level Explanatory Closure. It does it by using, as starting point, the same explanatory resources of OSE but reorganising their content in a way that would be compatible with the model presented throughout this paper. 
	In other terms, YAI4Hu uses as explanandum the same resources used by OSE, but it makes them reachable only via the main primitive actions described in section \ref{sec:proof_of_concept}, reorganising information accordingly.
	
	The third system is a \textit{How-Why Narrator} (HWN, in short), another type of One-Size-Fits-All explanation, and it is YAI4Hu but without \textit{Open Question Answering} and with \textit{Aspect Overviewing} only for \quotes{how} and \quotes{why} explanations (i.e. \quotes{what} or \quotes{who} explanations are not considered).
	
	\subsection{User-Study: Questionnaires and Participants} \label{sec:user_study}
	In order to verify hypothesis \ref{hyp:main}, we designed a user-study involving the 2 explananda and the 3 explanatory approaches.
	
	\edit{We recruited 68 different and anonymous participants among the students of our university. These students came from a few different courses of study:
	\begin{itemize}
		\item Bachelor Degree in Computer Science, age in $[19,23]$.
		\item Bachelor Degree in Management for Informatics, age in $[19,23]$.
		\item Master Degree in Digital Humanities, age in $[21,25]$.
		\item Master Degree in Artificial Intelligence, age in $[21,25]$.
	\end{itemize}}
	Only the master degrees are international, with students from different countries and English teachings.
	
	To measure effectiveness and efficiency we designed two domain-specific quizzes (one per explanandum), covering three different archetypes: why, how and what.
	Each question in the quiz represents an informative goal for one or more users. 
	Being impossible and unfeasible to identify all the possible questions a real user would ask to reach its goal, we decided to select a few representative ones for the sake of the study.
	
	We picked different types of questions, with different archetypes and complexities, using as reference for each explanandum the main user goals discussed in section \ref{sec:explananda}.
	In fact, both the heart disease predictor and the credit approval system have different but well-defined purposes.
	Most importantly, many of the questions have been selected so that:
	\begin{itemize}
		\item Providing the correct answers would require the exploration of at least 2 or 3 different \textit{Aspect Overviews}, in HWN and YAI4Hu.
		\item The answers reachable via \textit{Open Question Answering} (in YAI4Hu) are not always as accurate as required (with the correct ones not ranked first) or are wrong (questions 1 and 6 of the \textit{credit approval system} quiz and questions 1, 2 and 3 of the \textit{heart disease predictor} quiz).
	\end{itemize}
	For each question we selected 4 to 8 different plausible answers of which only one was (the most) correct. One of the (wrong) answers was always \quotes{I don't know}.
	
	The heart disease predictor is designed to facilitate a responder predicting the likelihood of heart disease of a caller, suggesting the next concrete actions to take (i.e. a test, a new habit, etc.) to treat or avoid the disease, in accordance with the biological parameters.
	\edit{The questions we selected for the quiz on the heart disease predictor are shown in Table \ref{tab:hd_quiz}.}
	\begin{table}
		\centering
		\caption{\edit{
			\textbf{Quiz - Heart Disease Predictor}: \textit{questions}, \textit{type} and \textit{steps} are shown. 
			\textit{Type} indicates which interrogative particle is representative of the question. 
			\textit{Steps} is the minimum number of steps (in terms of links to click, overviews to open and/or questions to pose) required by each explanatory tool. 
			Negative \textit{steps} means that the correct answer cannot be found, while 0 \textit{steps} means that the answer is immediately available without clicking on any link.
			On the other hand, \quotes{no OQA} means that Open Question Answering does not answer correctly to the question.
		}}
		\label{tab:hd_quiz}
		\resizebox{\linewidth}{!}{
			\arrayrulecolor{black}
			\begin{tabular}{|p{0.6\linewidth}|l|l|l|l|} 
				\hline
				\rowcolor[rgb]{0.937,0.937,0.937} {\cellcolor[rgb]{0.937,0.937,0.937}}                                                     & {\cellcolor[rgb]{0.937,0.937,0.937}}                                & \multicolumn{3}{l|}{\textbf{Steps}}                                        \\ 
				\hhline{|>{\arrayrulecolor[rgb]{0.937,0.937,0.937}}-->{\arrayrulecolor{black}}---|}
				\rowcolor[rgb]{0.937,0.937,0.937} \multirow{-2}{*}{{\cellcolor[rgb]{0.937,0.937,0.937}}\textbf{Question}}                  & \multirow{-2}{*}{{\cellcolor[rgb]{0.937,0.937,0.937}}\textbf{Type}} & \textbf{\textbf{OSE~}} & \textbf{\textbf{HWN}} & \textbf{\textbf{YAI4Hu}}  \\ 
				\hline
				What are the most important factors leading that patient to a medium risk of heart disease?                                & what, why                                                           & 0                      & 0                     & 0 (no OQA)                         \\ 
				\hline
				What is the easiest thing that the patient could actually do to change his heart disease risk from medium to low?          & what, how                                                           & 0                      & 0                     & 0 (no OQA)                         \\ 
				\hline
				According to the predictor, what level of serum cholesterol is needed to shift the heart disease risk from medium to high? & what, how                                                           & 0                      & 0                     & 0 (no OQA)                         \\ 
				\hline
				How could the patient avoid raising bad cholesterol, preventing his heart disease risk to shift from medium to high?       & how                                                                 & 1                      & 2                     & 2                         \\ 
				\hline
				What kind of tests can be done to measure bad cholesterol levels in the blood?                                             & what, how                                                           & 1                      & -1                    & 1                         \\ 
				\hline
				What are the risks of high cholesterol?                                                                                    & what, why-not                                                       & 1                      & 2                     & 1                         \\ 
				\hline
				What is LDL?                                                                                                               & what                                                                & 1                      & 2                     & 1                         \\ 
				\hline
				What is Serum Cholestrol?                                                                                                  & what                                                                & 1                      & 1                     & 1                         \\ 
				\hline
				What types of chest pain are typical of heart disease?                                                                     & what, how                                                           & 1                      & 1                     & 1                         \\ 
				\hline
				What is the most common type of heart disease in the USA?                                                                  & what                                                                & 1                      & 1                     & 1                         \\ 
				\hline
				What are the causes of angina?                                                                                             & what, why                                                           & 1                      & 2                     & 1                         \\ 
				\hline
				What kind of chest pain do you feel with angina?                                                                           & what, how                                                           & 1                      & 1                     & 1                         \\ 
				\hline
				What are the effects of high blood pressure?                                                                               & what, why-not                                                       & 1                      & 1                     & 1                         \\ 
				\hline
				What are the symptoms of high blood pressure?                                                                              & what, why                                                           & 1                      & 1                     & 1                         \\ 
				\hline
				What are the effects of smoking to the cardiovascular system?                                                              & what, why-not                                                       & 1                      & 3                     & 1                         \\ 
				\hline
				How can the patient increase his heart rate?                                                                               & how                                                                 & 1                      & 3                     & 1                         \\ 
				\hline
				How can the patient try to prevent a stroke?                                                                               & how                                                                 & 1                      & 3                     & 2                         \\ 
				\hline
				What is a Thallium stress test?                                                                                            & what, why                                                           & 1                      &  3                    & 1                         \\
				\hline
			\end{tabular}
		}
	\end{table}
	Interestingly, many questions are polyvalent in the sense that they can be rewritten using different archetypes. For example the question \quotes{Why, in terms of factor, does that patient have a medium risk of heart disease?} can be rewritten as \quotes{What are the most important factors leading that patient to a medium risk of heart disease?}, or the question \quotes{How can an account become delinquent?} in \quotes{Why does an account become delinquent?}.
	
	The credit approval system is designed to help an applicant (i.e. Mary) to understand the results of its loan application and how to concretely change them, what to do to get the loan accepted instead of denied. We believe that in this context, a real user-centred system should answer to more than the question \quotes{What are the main factors responsible for the rejection?}.
	
	\edit{The questions we selected for the quiz on the credit approval system are shown in Table \ref{tab:ca_quiz}.}
	\begin{table}
		\centering
		\caption{\edit{
			\textbf{Quiz - Credit Approval System}: \textit{questions}, \textit{type} and \textit{steps} are shown. \textit{Type} indicates which interrogative particle is representative of the question. 
			\textit{Steps} is the minimum number of steps (in terms of links to click, overviews to open and/or questions to pose) required by each explanatory tool.  
			Negative \textit{steps} means that the correct answer cannot be found, while 0 \textit{steps} means that the answer is immediately available without clicking on any link. 
			On the other hand, \quotes{no OQA} means that Open Question Answering does not answer correctly to the question.
		}}
		\label{tab:ca_quiz}
		\resizebox{\linewidth}{!}{
			\arrayrulecolor{black}
			\begin{tabular}{|p{0.6\linewidth}|l|l|l|l|} 
				\hline
				\rowcolor[rgb]{0.937,0.937,0.937} {\cellcolor[rgb]{0.937,0.937,0.937}}                                                                                           & {\cellcolor[rgb]{0.937,0.937,0.937}}                                & \multicolumn{3}{l|}{\textbf{Steps}}                                        \\ 
				\hhline{|>{\arrayrulecolor[rgb]{0.937,0.937,0.937}}-->{\arrayrulecolor{black}}---|}
				\rowcolor[rgb]{0.937,0.937,0.937} \multirow{-2}{*}{{\cellcolor[rgb]{0.937,0.937,0.937}}\textbf{Question}}                                                        & \multirow{-2}{*}{{\cellcolor[rgb]{0.937,0.937,0.937}}\textbf{Type}} & \textbf{\textbf{OSE~}} & \textbf{\textbf{HWN}} & \textbf{\textbf{YAI4Hu}}  \\ 
				\hline
				What did the Credit Approval System decide for Mary's application?                                                                                               & what, how                                                           & 0                      & 0                     & 0                         \\ 
				\hline
				What is an inquiry (in this context)?                                                                                                                            & what                                                                & 1                      & 1                     & 1                         \\ 
				\hline
				What type of inquiries can affect Mary's score, the hard or the soft ones?                                                                                       & what, how                                                           & 1                      & 1                     & 1                         \\ 
				\hline
				What is an example of hard inquiry?                                                                                                                              & what                                                                & 1                      & -1                    & 1                         \\ 
				\hline
				How can an account become delinquent?                                                                                                                            & how, why                                                            & 1                      & 1                     & 1                         \\ 
				\hline
				Which specific process was used by the Bank to automatically decide whether to assign the loan?                                                                  & what, how                                                           & 0                      & 0                     & 0 (no OQA)                         \\ 
				\hline
				What are the known issues of the specific technology used by the Bank (to automatically predict Mary's risk performance and to suggest avenues for improvement)? & what, why                                                           & 1                      & 1                     & 1 (no OQA)           \\
				\hline
			\end{tabular}
		}
	\end{table}
	Now, it is important to note that the last two questions are about the specific technology used by the system. In fact, in this specific context, the data subject (the loan applicant) should be aware of the technological limitations and issues of the automated decision maker (the credit approval system), as suggested by the \ac{GDPR} and subsequent works including \cite{sovrano2020modelling}.
	
	We tried to keep the size of the two quizzes proportional to the complexity and richness of the explananda. Intuitively, the heart disease predictor is a much more complex explanandum with many more resources and questions to answer.
	\edit{Furthermore, we expect that for answering some of the questions (i.e. the first one of the Credit Approval System) it is sufficient to read the initial explanations provided by the systems. 
	Therefore, people failing to answer at least one question is likely to be answering (more or less) randomly/nonsensically, paying no attention to the task. 
	This is why we decided to use the number of correct answers as \textit{attention check}, discarding all participants with less than $1$ correct answer per quiz.}
	
	\edit{Each participant was asked to test both the explananda (starting from the credit approval system, the simplest one) but it was randomly allocated to test only one of the three explanatory approaches. In other terms, it was a between-subjects experimental design, so that every participant was assigned to test only one single explanatory tool (either OSE, HWN or YAI4Hu) and not multiple ones. 
	Furthermore, all the participants were asked to complete (in English) a quiz and a \ac{SUS} questionnaire per explanandum, and to optionally provide some qualitative feedback in the form of a comment. 
	Despite this, some participants refused to test the heart disease predictor because too burdensome in terms of minimum time required to complete the quiz.}
	Participants were told that completing the questionnaire (on both the explananda) would have taken an average time that varies from 10 to 25 minutes, and to use a desktop/laptop because the explanatory tools were not designed for touchscreens.
	They were also informed, in a simple and very concise way, that the goal of the survey was to understand which explanatory mechanism (among many) is the best one, without going into further details. Therefore they did know that other versions of the explanatory tool were available and that each other user may have received a different one.
	
	Our test evaluated effectiveness and satisfaction only on people with a normal \ac{NCS}, across a number of tasks meant to put the main archetypical questions in play. 
	The \ac{NCS} \cite{cacioppo1982need,lins2020very} is a user characteristic that refers to the user’s tendency to engage in and enjoy thinking. \ac{NCS} has become influential across social and medical sciences, and it is not new to the human-computer interaction community either \cite{millecamp2019explain}.
	According to \citeauthor{cacioppo1982need}\cite{cacioppo1982need}, \ac{NCS} can be measured through a specific questionnaire of 18 items, which responses are given on a 5-point scale (1 = extremely uncharacteristic of the user; 5 = extremely characteristic of user).
	In \citeyear{lins2020very}, \citeauthor{lins2020very}\cite{lins2020very} proposed a simplified version of the original questionnaire, called NCS-6 and with only 6 items instead of 18.
	
	\ac{NCS} is interesting to consider for our purposes, because the usability of an explanatory tool may be significantly different for people with a low, normal or high \ac{NCS}. 
	In fact, it is reasonable to assume that only the most dedicated and focussed users (those with a high \ac{NCS}) can handle (also with satisfaction) the effort to search in a One-Size-Fits-All Exhaustive Explanatory Closure.
	On the other end, users with a too low \ac{NCS} may be more prone to avoid any (also minimally) challenging cognitive task, especially if it involves understanding a complex-enough explanandum. Therefore users with low \ac{NCS} may be not satisfied at all of any possible explanatory tool, just because the underlying task makes them spend more than a few minutes.
	For these reasons we believe that it is important to test the usability of a user-centred explanatory tool on people with a normal \ac{NCS}, as we did.
	
	In order to understand whether a person has a \quotes{normal}\footnote{\edit{The word \quotes{normal} here does not indicate the normal distribution but instead the fact that the scores within the inter-quartile range are not extreme scores.}} \ac{NCS} we have to collect enough scores\footnote{In our case we were able to collect 50 \ac{NCS}.} and compute their interquartile range. 
	\ac{NCS} scores lying within the interquartile range are said to be \quotes{normal}, because the interquartile is the range of scores that are not too high nor too low.
	The interquartile range is meaningful when no assumption can be done on the distribution of scores across the population of users participating to the study. 
	
	\edit{For the credit approval system (CA) we got 68 participants, as shown in Table \ref{tab:participants}:}
	\begin{itemize}
		\item \textbf{OSE}: 21 participants, all passed the attention check but only 19 took the NCS-6 test.
		\item \textbf{HWN}: 18 participants, all passed the attention check but only 15 took the NCS-6 test.
		\item \textbf{YAI4Hu}: 29 participants, but 3 did not pass the attention check and of the others only 20 took the NCS-6 test.
	\end{itemize}
	\edit{For the heart disease predictor (HD) we got 55 participants, as shown in Table \ref{tab:participants}:}
	\begin{itemize}
		\item \textbf{OSE}: 17 participants, all passed the attention check but only 16 took the NCS-6 test.
		\item \textbf{HWN}: 17 participants, all passed the attention check but only 12 took the NCS-6 test.
		\item \textbf{YAI4Hu}: 21 participants, but 3 did not pass the attention check and of the others only 12 took the NCS-6 test.
	\end{itemize}

	\begin{table}
		\centering
		\caption{\edit{
			\textbf{User Study - Participants}: for both HD and CA and for each explanatory approach (OSE, HWN, YAI4Hu), this table shows the number of participants adhering to the user-study. The first column (\quotes{Respondents}) shows the total number of respondents. The second column (\quotes{Check}) shows only the number of respondents that passed the \textit{attention check}. The third column (\quotes{Check+NCS}) shows only the number of respondents that passed the \textit{attention check} and completed the NCS-6 questionnaire. Box-plots \ref{fig:ncs_boxplot}, \ref{fig:poc_results_boxplot} and \ref{fig:poc_results_boxplot_any_ncs} are considering only the respondents of the third column.
		}}
		\label{tab:participants}
			\arrayrulecolor{black}
			\begin{tabular}{|l|l|l|l|l|} 
				\hhline{~~---|}
				\multicolumn{2}{l|}{}                                                                                 & {\cellcolor[rgb]{0.937,0.937,0.937}}Respondents & {\cellcolor[rgb]{0.937,0.937,0.937}}Check & {\cellcolor[rgb]{0.937,0.937,0.937}}Check+NCS  \\ 
				\hline
				{\cellcolor[rgb]{0.937,0.937,0.937}}                     & {\cellcolor[rgb]{0.937,0.937,0.937}}OSE    & 21                                            & 21                                          & 19                                               \\ 
				\hhline{|>{\arrayrulecolor[rgb]{0.937,0.937,0.937}}->{\arrayrulecolor{black}}----|}
				{\cellcolor[rgb]{0.937,0.937,0.937}}                     & {\cellcolor[rgb]{0.937,0.937,0.937}}HWN    & 18                                            & 18                                          & 15                                               \\ 
				\hhline{|>{\arrayrulecolor[rgb]{0.937,0.937,0.937}}->{\arrayrulecolor{black}}----|}
				\multirow{-3}{*}{{\cellcolor[rgb]{0.937,0.937,0.937}}CA} & {\cellcolor[rgb]{0.937,0.937,0.937}}YAI4Hu & 29                                            & 26                                          & 20                                               \\ 
				\hhline{|=====|}
				{\cellcolor[rgb]{0.937,0.937,0.937}}                     & {\cellcolor[rgb]{0.937,0.937,0.937}}OSE    & 17                                            & 17                                          & 16                                               \\ 
				\hhline{|>{\arrayrulecolor[rgb]{0.937,0.937,0.937}}->{\arrayrulecolor{black}}----|}
				{\cellcolor[rgb]{0.937,0.937,0.937}}                     & {\cellcolor[rgb]{0.937,0.937,0.937}}HWN    & 17                                            & 17                                          & 12                                               \\ 
				\hhline{|>{\arrayrulecolor[rgb]{0.937,0.937,0.937}}->{\arrayrulecolor{black}}----|}
				\multirow{-3}{*}{{\cellcolor[rgb]{0.937,0.937,0.937}}HD} & {\cellcolor[rgb]{0.937,0.937,0.937}}YAI4Hu & 21                                            & 18                                          & 12                                               \\
				\hline
			\end{tabular}
	\end{table}
	
	At the end we had 54 valid participants taking the NCS-6 test for CA and only 40 of them for HD. 
	The \ac{NCS} score is computed by summing the given points (from 1 to 5 for questions 1,2,5 and 6; from -5 to -1 for questions 3 and 4) for each item of the NCS-6 questionnaire. 
	The resulting \ac{NCS} median score was $7$ with a lower quartile of $5$ and a upper quartile of $11$. Therefore participants with a \quotes{normal} \ac{NCS} score $s$ were those with $5 \leq s \leq 11$. 
	
\begin{figure}[!htb]
		\centering
		\includegraphics[width=.9\columnwidth]{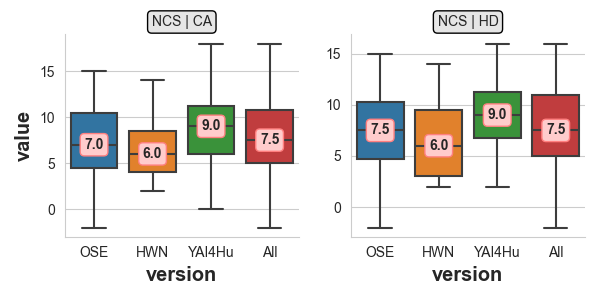}
		\caption{\textbf{\ac{NCS} scores} of those participants that passed the attention check. Results are shown in the form of box plots (25th, 50th, 75th percentile, and whiskers covering all data and outliers). The numerical value of medians is shown inside pink boxes. Results for OSE are in blue, for HWN are in orange, for YAI4Hu are in green, for all the explanatory tools are in red.}
		\label{fig:ncs_boxplot}
	\end{figure}

	The mean \ac{NCS} was $7.49$ suggesting that the collected \ac{NCS} scores are normally distributed. The box-plot of the valid \ac{NCS} scores for CA and HD is shown in figure \ref{fig:ncs_boxplot}.
	
	For answering the effectiveness quizzes, participants were repeatedly asked to use only the information reachable from within the systems (i.e. by following the external hyper-links in there). In other terms, they were clearly instructed to not use Google or other external tools for answering.
	Participants were also:
	\begin{itemize}
		\item Instructed to click on \quotes{I don't know} in case they do not know an answer.
		\item Informed that there is only one correct answer for each question and when multiple answers seem to be reasonably correct, only the most precise is considered to be the correct one.
		\item Noticed when a wrong answer was given, showing them the correct one, in order to make them aware of their success or failure in reaching a goal. Questions were shown in order, one by one, separately, and answers were randomly shuffled.
	\end{itemize}

	At the end of the effectiveness quiz the answers were automatically scored as correct (score 1) or not (score 0). For example for the question \quotes{What did the Credit Approval System decide for Mary's application?} the correct answer is \quotes{It was rejected} and wrong answers are \quotes{Nothing} or \quotes{I don't know}.
	
	\section{Results Discussion} \label{sec:results}
	\begin{figure}[!htb]
		\centering
		\includegraphics[width=.9\columnwidth]{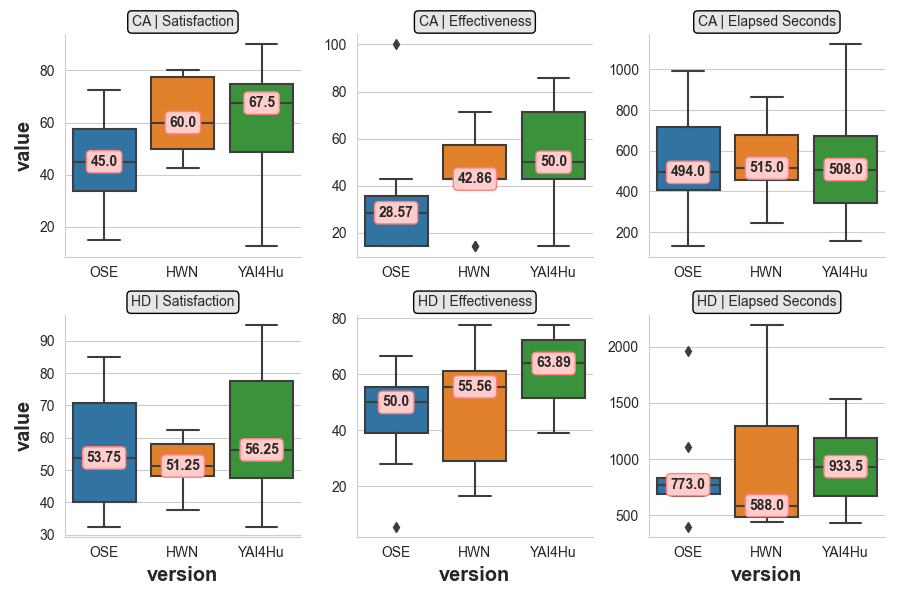}
		\caption{\textbf{Usability for Participants with a \quotes{Normal} NCS}: Results are shown in the form of box plots (25th, 50th, 75th percentile, and whiskers covering all data and outliers). The numerical value of medians is shown inside pink boxes. Results for OSE are in blue, for HWN are in orange and for YAI4Hu are in green.}
		\label{fig:poc_results_boxplot}
	\end{figure}
	
	We defined our results measures as:
	\begin{inparadesc}
		\item \textbf{Satisfaction},
		\item \textbf{Effectiveness},
		\item and \textbf{Elapsed Seconds} (that are inversely proportional to Efficiency).
	\end{inparadesc}
	
	In Figure \ref{fig:poc_results_boxplot} we show the resulting box-plots, for every given measure, on all the explananda and explanatory approaches, and on participants with a \quotes{normal} \ac{NCS} (as discussed in section\ref{sec:user_study}). 
	According to these box-plots, results seem to indicate that hypothesis \ref{hyp:main} is correct.
	In fact, we can see from figure \ref{fig:poc_results_boxplot} a clear trend of increasing effectiveness and satisfaction from OSE to YAI4Hu, aligned to our expectation that more \textit{illocution} and \textit{goal-orientedness} imply more usability.
	
	\edit{Interestingly and differently from the results shown in \cite{sovrano2021philosophy}, here we have that satisfaction increases with effectiveness almost proportionally.
	We believe this is because in this experiment we made sure that participants' objective was to complete the quizzes with the best score possible, by not paying or rewarding them\footnote{\edit{If participants only participated in the study because they would get paid/rewarded, their goal would be to get money as fast as possible and not to complete the quizzes with a good score.}} and by explicitly and immediately informing them when failing or succeeding.}
	
	
	Interestingly, the experiments show that \textit{illocution} can lead to an important increment in effectiveness on both the considered explananda, and that this increment can be slightly improved with more \textit{goal-orientedness}. 
	\edit{On the other side, it seems that satisfaction is more driven by \textit{goal-orientedness} than \textit{illocution}\footnote{A more adequate statistical analysis would be needed to have more certainty about this phenomenon, to understand if there is a strong correlation (Pearson or Spearman) or even causality (mixed model) between satisfaction, \textit{goal orientation} and \textit{illocution}.}. 
	These insights are also supported by the qualitative feedbacks provided by the participants that were:
	\begin{itemize}
		\item \textbf{Overall negative for OSE}, i.e. \quotes{I did not really understand the purpose of the website for this quiz, as i did not feel it helped anything. If the point was to use the available links at the site, there were too many of them, so that it was no longer useful.}
		\item \textbf{Overall neutral for HWN}. Most of the comments were like \quotes{I have no comment}, but for a few negative ones, i.e. \quotes{Too long, too difficult, strange way to ask the question... it wasn't very clear!}.
		\item \textbf{Overall positive for YAI4Hu}, i.e. \quotes{This time, the accuracy was surprisingly great: most of the times, the correct answer was the first to be given. However, in a couple of cases the answer wasn't even among the ones given (and the system still counted them as "sufficient"): I noticed that this happens especially with more general questions, such as "what is ...", and therefore had to click on the name to know the right answer, while more specific questions (such has "what causes..." or "who suffers most...") were easier for the system to find}. Though, some suggestions for improvements were also given, i.e. \quotes{The given information for each answer was a lot, and not always the answer I was looking for was among the first; also, there could have been more possible answers with for the same question, but separated in the list} .
	\end{itemize}}

	\edit{The results shown in Figure \ref{fig:poc_results_boxplot} indicate that the distribution of scores is skewed, with medians that are usually closer to one of the other quartiles.
	Therefore, due to the limited number of samples, we choose to not make assumptions of parametrisation in the data\footnote{Anonymised data is available at \url{https://github.com/Francesco-Sovrano/YAI4Hu}, for reproducibility purposes.} collected through the user-study, this forced us to rely on non-parametric tests.}
	To fully verify the hypothesis, discarding the possibility that the outcomes are the result of luck, we performed a few one-sided Mann-Whitney U-tests (MW; a non-parametric version of the t-test for independent samples) and Kruskal-Wallis H-tests (KW; a non-parametric version of ANOVA) on the global (between-subjects) scores.
	
	\edit{For the Credit Approval System (CA) the results (without Bonferroni correction) are:}
	\begin{itemize}
		\item \textbf{Effectiveness score}: OSE's effectiveness is lower than HWN's according to MW (U=28, p=0.050). Furthermore, data also show that OSE's Effectiveness is significantly\footnote{Assuming $p < 0.05$ is enough for asserting significance.} lower than YAI4Hu's (MW: U=28, p=0.009).
		\item \textbf{Satisfaction score}: OSE's is significantly lower than HWN's (MW: U=25, p=0.033) and lower than YAI4Hu's (MW: U=39.5, p=0.054).
		\item \textbf{Elapsed Seconds}: OSE's are comparable to HWN's according to KW (H=0.11, p=0.73), but nothing certain can be said for YAI4Hu's (KW: H=0.06, p=0.80) compared to OSE's.
	\end{itemize}
	
	\edit{For the Heart Disease Predictor (HD) the results (without Bonferroni correction) are:}
	\begin{itemize}
		\item \textbf{Effectiveness score}: OSE's effectiveness is significantly lower than YAI4Hu's (MW: U=19, p=0.032), while nothing certain can be said with respect to HWN's (MW: U=26, p=0.35).
		\item \textbf{Satisfaction score}: nothing certain can be said about HWN's (KW: H=0.14, p=0.70) and YAI4Hu's (KW: H=0.38, p=0.53), compared to OSE's.
		\item \textbf{Elapsed Seconds}: nothing certain can be said for both HWN and YAI4Hu, compared to OSE's.
	\end{itemize}

	\edit{Considering that we are doing 3 multiple comparisons per score (effectiveness, satisfaction, seconds) with MW/KW, the chances of having a comparison that falsely results as expected increase.
	Some statistical tools that are used in this case, to reduce the chance of a type I error (false positive), are: 
	\begin{inparadesc}
		\item the Bonferroni correction,
		\item the Holm–Bonferroni method,
		\item or the Dunn–Šidák correction.
	\end{inparadesc}
	Though, these tools are known to concretely increase type II errors (false negatives) \cite{armstrong2014use}.
	Regardless, if we would use a Bonferroni or Dunn correction, to adjust for 3 multiple comparisons per score, then the minimum p-value for claiming a statistically significant result would not be $0.05$ but instead something close to $0.016$.
	Therefore, with these corrections, only one claim of statistical significance would still hold: in CA, OSE's effectiveness is significantly lower than YAI4Hu's (MW: U=28, p=0.009).}
	
	Anyway, the obtained results highlight a good correlation between objective (effectiveness) and subjective (satisfaction) metrics in both CA and HD, even if it is more evident in CA and very smooth in HD.
	We believe that this difference between CA's results and HD's is due to a couple of factors. 
	
	The first factor is that HD's quiz is much harder, considering that none of the participants was able to achieve an effectiveness score greater than 80\%.
	\edit{Indeed, the average number of steps that are minimally required, to reach the piece of information containing an answer, is higher in HD than CA, as show in Table \ref{tab:hd_quiz}.}
	This first factor may suggest that the satisfaction for an explanatory process is affected by the intrinsic complexity of the explanandum, in a different way from effectiveness.
	
	The second factor is that the 2nd information level of HD's OSE is mainly composed by web-pages from the website of the \textit{U.S. Centers for Disease Control and Prevention}\footnote{\url{https://www.cdc.gov}}, that usually organise its information as a FAQ. So that every page is practically a list of a few specific questions about a few aspects to explain, followed by a (usually) fairly long explanatory answer, making the content more usable. 
	This last factor should partly justify the fact that in HD there is little or no difference from OSE's satisfaction scores, in both HWN's and YAI4Hu's.
	
	\edit{Furthermore, the difference in usability between participants with \quotes{normal} \ac{NCS} and non \quotes{normal} \ac{NCS} can be seen by looking at the differences between figures \ref{fig:poc_results_boxplot} and \ref{fig:poc_results_boxplot_any_ncs}, showing the usability for all participants, regardless their \ac{NCS}. As hypothesised in section \ref{sec:user_study} we can see a drop in satisfaction for the more user-centred tools and a slight increment in OSE's effectiveness. 
	In fact only people with an high \ac{NCS} is usually satisfied and effective with extremely verbose explanatory contents as OSE's.
	\begin{figure}[!htb]
		\centering
		\includegraphics[width=.9\columnwidth]{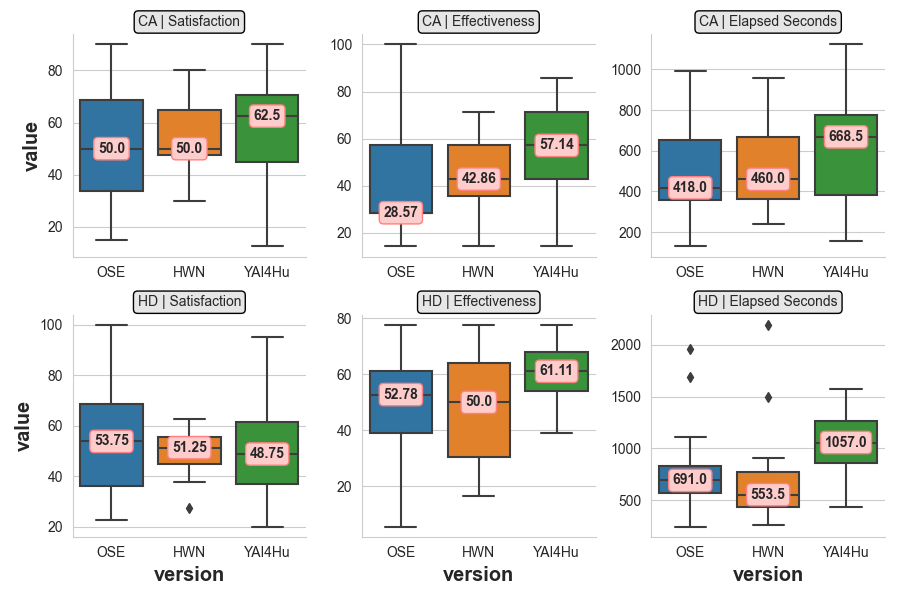}
		\caption{\textbf{Usability for All Participants, regardless their NCS}: Results are shown in the form of box plots (25th, 50th, 75th percentile, and whiskers covering all data and outliers). The numerical value of medians is shown inside pink boxes. Results for OSE are in blue, for HWN are in orange and for YAI4Hu are in green.}
		\label{fig:poc_results_boxplot_any_ncs}
	\end{figure}}

	\edit{Finally, we also investigate whether the increase in effectiveness over OSE is only due to the fact that YAI4Hu and HWN offer advanced mechanisms to easily navigate information.
	Surprisingly, the results show that YAI4Hu outperforms HWN and OSE also in all those questions that can be answered by simply reading the initial explanation (the few lines of text in the landing page), as shown in Figure \ref{fig:answers_in_initial_explanans_boxplot}.
	These questions are those that require 0 steps and are shown in Table \ref{tab:hd_quiz} and \ref{tab:ca_quiz}.
	\begin{figure}[!htb]
		\centering
		\includegraphics[width=.65\columnwidth]{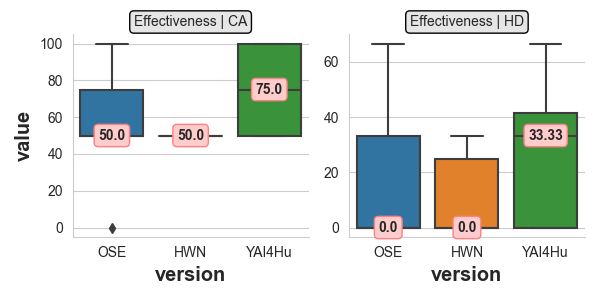}
		\caption{\edit{
			\textbf{Effectiveness scores of participants with a \quotes{Normal} NCS on the questions requiring 0 steps}: Results are shown in the form of box plots (25th, 50th, 75th percentile, and whiskers covering all data and outliers). 
			The numerical value of medians is shown inside pink boxes. 
			Results for OSE are in blue, for HWN are in orange and for YAI4Hu are in green. 
			The questions requiring 0 steps are shown in Table \ref{tab:hd_quiz} and \ref{tab:ca_quiz}.
		}}
		\label{fig:answers_in_initial_explanans_boxplot}
	\end{figure}}

	\edit{More in detail, we have statistically significant results for CA (even with a Bonferroni correction):
	\begin{itemize}
		\item \textbf{CA}: HWN's effectiveness is significantly lower than YAI4Hu's (MW: U=27, p=0.008) for questions requiring 0 steps.
		\item \textbf{HD}: HWN's effectiveness seems lower than YAI4Hu's (MW: U=15, p=0.11) for questions requiring 0 steps, but in this case results are not significant.
	\end{itemize}
	These results hold also if we consider the participants being discarded for not having passed the \textit{attention check} (with YAI4Hu).
	It could be that this difference between CA and HD is due to the fact that CA has one question requiring 0 steps that can also be answered via Open Question Answering, as shown in Table \ref{tab:ca_quiz}.
	However, in favour of our hypothesis, these results might suggest that effective explanations require more than merely showing to the user a sentence containing the precise answer.}
	
	\edit{Nonetheless, the limitations of this work are many. 
	First, it has only been evaluated with a small pool of students and this is clearly an issue that prevents us from making strong claims about the statistical evidence gathered during the user-study.
	Second, our evaluation of the explanatory mechanisms is entangled with that of the user interface, making hard to understand what are the main sources of usability issues.
	Third, the pipeline of algorithms relies on several heuristics and approximations that altogether might hinder the usability of the explanatory system. For example, the question-answer retrieval mechanism is far from perfect and in several occasions it fails to provide the best answer, as pointed out by several users.}
	
	\edit{\section{Conclusions and Future Work}} \label{sec:conclusions}
	In this paper, we proposed a new method for explanations in \ac{AI} and, consequently, a tool to test its expressive power within a user interface.
	Being interested in modelling an explanatory process for producing user-centric explanatory software, and in quantifying the difference it bears in terms of effectiveness with respect to non-pragmatic approaches, our solution drawn from state-of-the-art philosophical theories of explanation.
	
	Among the few philosophical theories of explanation, we identified the one that, we believe, is mostly convertible into a practical model for user-centric explanatory software: Achinstein's.
	But Achinstein's is an abstract illocutionary theory of explanation, therefore we proposed a way to concretely implement \textit{illocution} as the act of \textit{pertinently answering} implicit questions (i.e. Why? What for? How? When? etc..).
	
	What we showed is that an abstract philosophical theory of explanations can be beneficially implemented into a concrete software, as a question answering process.
	In fact, through the identification of a minimal set of archetypal questions, it is possible to obtain a generator of explanatory overviews generic enough to be able to significantly ease the acquisition of knowledge, regardless of the specific user but depending instead on a fairly broad category of selected users, thus resulting in a user-centred explanatory tool that is more effective than its non-pragmatic counterpart on the same explanandum.
	
	Our hypothesis was that, given an arbitrary explanatory process, increasing its \textit{goal-orientedness} and \textit{illocutionary power} results in the generation of more usable explanations.
	In other terms, we believe that the usability (as per ISO 9241-210) of an explanatory process depends on its ability to be \textit{illocutionary} and \textit{goal-oriented}.
	
	In order to test our hypothesis, we had to invent a new pipeline of \ac{AI} algorithms (briefly summarised in figure \ref{fig:yai_diagram}) and run a user-study on it. 
	This pipeline was able to organize the information contained in non-structured documents written in natural language (e.g. web pages, pdf, etc..), allowing efficient information clustering, according to a set of archetypal questions, aiming to build a sufficiently rich and effectively explorable \acf{ES} for the automated generation of user-centred explanations. 
	
	We tested our hypothesis on two \ac{XAI}-powered systems for credit approval and heart disease prediction, comparing different explanatory approaches when varying the \textit{illocutionary power} and \textit{goal-orientedness}.
	The results of the user-study, involving more than 60 participants, showed that our proposed solutions produced statistically relevant improvements on effectiveness (hence a p-value lower than 0.05) over the baseline. This gives evidence in favour of our hypothesis, also considering that the increment in effectiveness is visibly aligned with an increment of satisfaction.
	
	\edit{Though, we envisage that further analysis in a different context and with different participants might be required to verify the main hypothesis and evaluate the explanatory processes more thoroughly.
	For example, we think that another context of application of our technology could be that of artificial intelligence and education.
	Indeed, not surprisingly, many would argue that explanations are one of the main artefacts through which humans understand reality and learn to solve complex problems \cite{berland2009making}.
	Therefore, \textit{explaining} is not only central to \ac{XAI} but also to education and artificial intelligence, and these are two contexts where our technology and our understanding of explanations could be of utmost importance.}
	
	\edit{Hence, we are currently investigating on how to apply our extension of Achinstein's theory to explain external (formal) regulations to Reinforcement Learning agents.
	So far, the results look promising, as shown in \cite{sovrano2021explanation}, suggesting that our model of an explanatory process might be generic enough to work with both human and artificial intelligence.
	Besides, another future direction of work is that of using the model for the production of personalised educational contents, integrating it with existing technology for Knowledge Tracing \cite{thaker2018dynamic}.
	This might give us the opportunity to test the technology with a larger pool of users (i.e. a whole classroom) over a longer period of time (i.e. a semester), collecting more data about their behaviour.
	In fact, during our user-study we did not study the behaviour of participants (i.e. number of clicks, frequency of scrolling, etc..) and this could give vital insights about the underlying explanatory processes.}
	
	\edit{Overall, we believe that this work has somehow the potential to stress even further that more emphasis in the research of explainable and explanatory AI \cite{sovrano2020making} should be put on a proper understanding of what constitutes the act of explaining.
	This is why we re-elaborated several ideas coming from Achinstein's theory of explanations.
	Indeed, we argue that a large portion of XAI literature definitely puts an emphasis on the part of explaining that requires \quotes{pertinently and deliberately answering}. 
	Though, the critical link to usability is that explaining requires \quotes{illocution}, and that is answering also to the \quotes{implicit questions} from the user.
	This is somehow neglected by work that treats explanations only as a product, independent of the explainee's goals or knowledge.}
	
	
	\bibliographystyle{ACM-Reference-Format}
	\bibliography{biblio}
	
	\appendix
	
	%
	%
	
	\begin{acronym}
		\acro{EU}{European Union}
		\acro{ADM}{Automated Decision-Making system}
		\acro{AI-HLEG}{High-Level Expert Group on Artificial Intelligence}
		\acro{AI}{Artificial Intelligence}
		\acro{XAI}{eXplainable AI}
		\acro{YAI}{explanatorY AI}
		\acro{HCI}{Human-Computer Interaction}
		\acro{RL}{Reinforcement Learning}
		\acro{EN}{Explanatory Narrative}
		\acro{ENs}{Explanatory Narratives}
		\acro{EP}{Explanatory Process}
		\acro{ES}{Explanatory Space}
		\acro{GDPR}{General Data Protection Regulation}
		\acro{ETTAI}{Explanatory Tool for Trustworthy AI}
		\acro{EI}{Explainable Information}
		\acro{XP}{eXplainable Processes}
		\acro{XD}{eXplainable Datasets}
		\acro{UI}{User Interface}
		\acro{RDF}{Resource Description Framework}
		\acro{AIX360}{AI Explainability 360}
		\acro{CEM}{Contrastive Explanations Method}
		\acro{UCET}{User-Centred Explanatory Tool}
		\acro{SET}{Static Explanatory Tool}
		\acro{KB}{Knowledge Base}
		\acro{TFIDF}{Term Frequency–Inverse Document Frequency}
		\acro{USE}{Universal Sentence Encoder}
		\acro{SUS}{System Usability Scale}
		\acro{EU}{European Union}
		\acro{KG}{Knowledge Graph}
		\acro{TFIDF}{Term Frequency–Inverse Document Frequency}
		\acro{OKE}{Open Knowledge Extraction}
		\acro{NLP}{Natural Language Processing}
		\acro{FCA}{Formal Concept Analysis}
		\acro{QA}{Question Answering}
		\acro{ODP}{Ontology Design Pattern}
		\acro{PIL}{International Private Law}
		\acro{NCS}{Need for Cognition Score}
	\end{acronym}
	
\end{document}